\documentclass[11pt]{article}

\usepackage[preprint]{acl}

\usepackage{times}
\usepackage{latexsym}

\usepackage[T1]{fontenc}

\usepackage[utf8]{inputenc}

\usepackage{microtype}

\usepackage{inconsolata}

\usepackage{graphicx}

\usepackage{booktabs} 
\usepackage{footnotehyper}
\usepackage{hyperref}
\usepackage{savefnmark}
\usepackage{tablefootnote}

\usepackage{algorithm}
\usepackage{algorithmic}
\usepackage{amsmath}
\usepackage{amssymb}
\usepackage{mathtools}
\usepackage{amsthm}
\usepackage{subcaption}
\usepackage{utfsym}
\usepackage{tcolorbox}
\usepackage{pifont}
\usepackage{makecell} 

\tcbuselibrary{breakable}  
\tcbuselibrary{skins}      
\definecolor{PineGreen}{HTML}{007C4A} 
\definecolor{OrangeRed}{HTML}{FF4500} 
\definecolor{mycolor}{gray}{0.92}    

\usepackage[capitalize,noabbrev]{cleveref}

\theoremstyle{plain}

\theoremstyle{definition}

\theoremstyle{remark}

\usepackage{dcolumn, array, booktabs}
\usepackage{multirow}
\usepackage{cuted}
\usepackage[table]{xcolor} 
\usepackage[dvipsnames]{xcolor} 
\definecolor{skyblue}{RGB}{210,235,255}
\definecolor{lightcoral}{RGB}{255,228,225}
\definecolor{lemonchiffon}{RGB}{255,250,205}

\definecolor{prompt}{RGB}{173, 216, 230}
\definecolor{prompt-frame}{RGB}{0, 51, 102}
\definecolor{prompt2}{RGB}{173, 216, 230}
\definecolor{prompt2-frame}{RGB}{0, 51, 102}
\definecolor{prompt3}{RGB}{173, 216, 230}
\definecolor{prompt3-frame}{RGB}{0, 51, 102}
\definecolor{prompt4}{RGB}{173, 216, 230}
\definecolor{prompt4-frame}{RGB}{0, 51, 102}
\definecolor{prompt5}{RGB}{173, 216, 230}
\definecolor{prompt5-frame}{RGB}{0, 51, 102}
\usepackage{fontawesome5}
\tcbset{
    prompt_func/.style={
        enhanced,
        breakable,
        colback=prompt,        
        colframe=prompt-frame,            
        fontupper=\normalsize,               
        boxrule=1pt,                       
        arc=4mm,                           
        left=1mm, right=1mm, top=1mm, bottom=1mm, 
        boxsep=4pt,                        
        before skip=10pt, after skip=10pt, 
        overlay={
            \node[anchor=north west, xshift=4pt, text=white] at (frame.north west) {\faDatabase};
        },
        title={~~~~~~~\textbf{Prompt used with GPT-4o to generate annotation for Element Grounding with functional setting}},
        coltitle=white,
        fonttitle=\bfseries\small
    }
}

\tcbset{
    full_content_generation/.style={
        enhanced,
        breakable,
        colback=prompt,        
        colframe=prompt-frame,            
        fontupper=\normalsize,               
        boxrule=1pt,                       
        arc=4mm,                           
        left=1mm, right=1mm, top=1mm, bottom=1mm, 
        boxsep=4pt,                        
        before skip=10pt, after skip=10pt, 
        title={\faDatabase~~~\textbf{Prompt: Full Content Generation}},
        coltitle=white,
        fonttitle=\bfseries\small
    }
}
\tcbset{
    filter_figures/.style={
        enhanced,
        breakable,
        colback=prompt,        
        colframe=prompt-frame,            
        fontupper=\normalsize,               
        boxrule=1pt,                       
        arc=4mm,                           
        left=1mm, right=1mm, top=1mm, bottom=1mm, 
        boxsep=4pt,                        
        before skip=10pt, after skip=10pt, 
        title={\faDatabase~~~\textbf{Prompt: Filter Figures}},
        coltitle=white,
        fonttitle=\bfseries\small
    }
}
\tcbset{
    section_generation/.style={
        enhanced,
        breakable,
        colback=prompt,        
        colframe=prompt-frame,            
        fontupper=\normalsize,               
        boxrule=1pt,                       
        arc=4mm,                           
        left=1mm, right=1mm, top=1mm, bottom=1mm, 
        boxsep=4pt,                        
        before skip=10pt, after skip=10pt, 
        title={\faDatabase~~~\textbf{Prompt: Section Generation}},
        coltitle=white,
        fonttitle=\bfseries\small
    }
}
\tcbset{
    html_generation/.style={
        enhanced,
        breakable,
        colback=prompt,        
        colframe=prompt-frame,            
        fontupper=\normalsize,               
        boxrule=1pt,                       
        arc=4mm,                           
        left=1mm, right=1mm, top=1mm, bottom=1mm, 
        boxsep=4pt,                        
        before skip=10pt, after skip=10pt, 
        title={\faDatabase~~~\textbf{Prompt: HTML Generation}},
        coltitle=white,
        fonttitle=\bfseries\small
    }
}
\tcbset{
    full_content_review/.style={
        enhanced,
        breakable,
        colback=prompt,        
        colframe=prompt-frame,            
        fontupper=\normalsize,               
        boxrule=1pt,                       
        arc=4mm,                           
        left=1mm, right=1mm, top=1mm, bottom=1mm, 
        boxsep=4pt,                        
        before skip=10pt, after skip=10pt, 
        title={\faDatabase~~~\textbf{Prompt: Full Content Review}},
        coltitle=white,
        fonttitle=\bfseries\small
    }
}
\tcbset{
    full_content_revise/.style={
        enhanced,
        breakable,
        colback=prompt,        
        colframe=prompt-frame,            
        fontupper=\normalsize,               
        boxrule=1pt,                       
        arc=4mm,                           
        left=1mm, right=1mm, top=1mm, bottom=1mm, 
        boxsep=4pt,                        
        before skip=10pt, after skip=10pt, 
        title={\faDatabase~~~\textbf{Prompt: Full Content Revise}},
        coltitle=white,
        fonttitle=\bfseries\small
    }
}
\tcbset{
    text_content_generation/.style={
        enhanced,
        breakable,
        colback=prompt,        
        colframe=prompt-frame,            
        fontupper=\normalsize,               
        boxrule=1pt,                       
        arc=4mm,                           
        left=1mm, right=1mm, top=1mm, bottom=1mm, 
        boxsep=4pt,                        
        before skip=10pt, after skip=10pt, 
        title={\faDatabase~~~\textbf{Prompt: Text Content Generation}},
        coltitle=white,
        fonttitle=\bfseries\small
    }
}
\tcbset{
    html_review/.style={
        enhanced,
        breakable,
        colback=prompt,        
        colframe=prompt-frame,            
        fontupper=\normalsize,               
        boxrule=1pt,                       
        arc=4mm,                           
        left=1mm, right=1mm, top=1mm, bottom=1mm, 
        boxsep=4pt,                        
        before skip=10pt, after skip=10pt, 
        title={\faDatabase~~~\textbf{Prompt: HTML Review}},
        coltitle=white,
        fonttitle=\bfseries\small
    }
}
\tcbset{
    html_revise/.style={
        enhanced,
        breakable,
        colback=prompt,        
        colframe=prompt-frame,            
        fontupper=\normalsize,               
        boxrule=1pt,                       
        arc=4mm,                           
        left=1mm, right=1mm, top=1mm, bottom=1mm, 
        boxsep=4pt,                        
        before skip=10pt, after skip=10pt, 
        title={\faDatabase~~~\textbf{Prompt: HTML Revise}},
        coltitle=white,
        fonttitle=\bfseries\small
    }
}
\tcbset{
    prompt_aethetic_element/.style={
        enhanced,
        breakable,
        colback=prompt,        
        colframe=prompt-frame,            
        fontupper=\normalsize,               
        boxrule=1pt,                       
        arc=4mm,                           
        left=1mm, right=1mm, top=1mm, bottom=1mm, 
        boxsep=4pt,                        
        before skip=10pt, after skip=10pt, 
        title={\faDatabase~~~\textbf{Prompt: Element Quality Judge}},
        coltitle=white,
        fonttitle=\bfseries\small
    }
}

\tcbset{
    prompt_aethetic_Layout/.style={
        enhanced,
        breakable,
        colback=prompt,        
        colframe=prompt-frame,            
        fontupper=\normalsize,               
        boxrule=1pt,                       
        arc=4mm,                           
        left=1mm, right=1mm, top=1mm, bottom=1mm, 
        boxsep=4pt,                        
        before skip=10pt, after skip=10pt, 
        title={\faDatabase~~~\textbf{Prompt: Layout Balance Judge}},
        coltitle=white,
        fonttitle=\bfseries\small
    }
}

\tcbset{prompt_aethetic_engagement/.style={
        enhanced,
        breakable,
        colback=prompt,        
        colframe=prompt-frame,            
        fontupper=\normalsize,               
        boxrule=1pt,                       
        arc=4mm,                           
        left=1mm, right=1mm, top=1mm, bottom=1mm, 
        boxsep=4pt,                        
        before skip=10pt, after skip=10pt, 
        title={\faDatabase~~~\textbf{Prompt: Engagement Judge}},
        coltitle=white,
        fonttitle=\bfseries\small
    }
}

\tcbset{prompt_info_clarity/.style={
        enhanced,
        breakable,
        colback=prompt,        
        colframe=prompt-frame,            
        fontupper=\normalsize,               
        boxrule=1pt,                       
        arc=4mm,                           
        left=1mm, right=1mm, top=1mm, bottom=1mm, 
        boxsep=4pt,                        
        before skip=10pt, after skip=10pt, 
        title={\faDatabase~~~\textbf{Prompt: Clarity Judge}},
        coltitle=white,
        fonttitle=\bfseries\small
    }
}

\tcbset{prompt_info_content/.style={
        enhanced,
        breakable,
        colback=prompt,        
        colframe=prompt-frame,            
        fontupper=\normalsize,               
        boxrule=1pt,                       
        arc=4mm,                           
        left=1mm, right=1mm, top=1mm, bottom=1mm, 
        boxsep=4pt,                        
        before skip=10pt, after skip=10pt, 
        title={\faDatabase~~~\textbf{Prompt: Content Completeness Judge}},
        coltitle=white,
        fonttitle=\bfseries\small
    }
}

\tcbset{prompt_info_logic/.style={
        enhanced,
        breakable,
        colback=prompt,        
        colframe=prompt-frame,            
        fontupper=\normalsize,               
        boxrule=1pt,                       
        arc=4mm,                           
        left=1mm, right=1mm, top=1mm, bottom=1mm, 
        boxsep=4pt,                        
        before skip=10pt, after skip=10pt, 
        title={\faDatabase~~~\textbf{Prompt: Logical Flow Judge}},
        coltitle=white,
        fonttitle=\bfseries\small
    }
}

\tcbset{prompt_verbatim_question/.style={
        enhanced,
        breakable,
        colback=prompt,        
        colframe=prompt-frame,            
        fontupper=\normalsize,               
        boxrule=1pt,                       
        arc=4mm,                           
        left=1mm, right=1mm, top=1mm, bottom=1mm, 
        boxsep=4pt,                        
        before skip=10pt, after skip=10pt, 
        title={\faDatabase~~~\textbf{Prompt: Generate Verbatim QA}},
        coltitle=white,
        fonttitle=\bfseries\small
    }
}

\tcbset{prompt_interpretive_question/.style={
        enhanced,
        breakable,
        colback=prompt,        
        colframe=prompt-frame,            
        fontupper=\normalsize,               
        boxrule=1pt,                       
        arc=4mm,                           
        left=1mm, right=1mm, top=1mm, bottom=1mm, 
        boxsep=4pt,                        
        before skip=10pt, after skip=10pt, 
        title={\faDatabase~~~\textbf{Prompt: Generate Interpretive QA}},
        coltitle=white,
        fonttitle=\bfseries\small
    }
}

\tcbset{prompt_answer_agent/.style={
        enhanced,
        breakable,
        colback=prompt,        
        colframe=prompt-frame,            
        fontupper=\normalsize,               
        boxrule=1pt,                       
        arc=4mm,                           
        left=1mm, right=1mm, top=1mm, bottom=1mm, 
        boxsep=4pt,                        
        before skip=10pt, after skip=10pt, 
        title={\faDatabase~~~\textbf{Prompt: Answer Questions}},
        coltitle=white,
        fonttitle=\bfseries\small
    }
}

\tcbset{prompt_4o_image/.style={
        enhanced,
        breakable,
        colback=prompt,        
        colframe=prompt-frame,            
        fontupper=\normalsize,               
        boxrule=1pt,                       
        arc=4mm,                           
        left=1mm, right=1mm, top=1mm, bottom=1mm, 
        boxsep=4pt,                        
        before skip=10pt, after skip=10pt, 
        title={\faDatabase~~~\textbf{Prompt: \texttt{4o-Image}}},
        coltitle=white,
        fonttitle=\bfseries\small
    }
}

\tcbset{prompt_owl_4o/.style={
        enhanced,
        breakable,
        colback=prompt,        
        colframe=prompt-frame,            
        fontupper=\normalsize,               
        boxrule=1pt,                       
        arc=4mm,                           
        left=1mm, right=1mm, top=1mm, bottom=1mm, 
        boxsep=4pt,                        
        before skip=10pt, after skip=10pt, 
        title={\faDatabase~~~\textbf{Prompt: \texttt{OWL-4o}}},
        coltitle=white,
        fonttitle=\bfseries\small
    }
}

\tcbset{prompt_4o_html/.style={
        enhanced,
        breakable,
        colback=prompt,        
        colframe=prompt-frame,            
        fontupper=\normalsize,               
        boxrule=1pt,                       
        arc=4mm,                           
        left=1mm, right=1mm, top=1mm, bottom=1mm, 
        boxsep=4pt,                        
        before skip=10pt, after skip=10pt, 
        title={\faDatabase~~~\textbf{Prompt: \texttt{4o-HTML}}},
        coltitle=white,
        fonttitle=\bfseries\small
    }
}

\tcbset{prompt_parser_summarizer/.style={
        enhanced,
        breakable,
        colback=prompt,        
        colframe=prompt-frame,            
        fontupper=\normalsize,               
        boxrule=1pt,                       
        arc=4mm,                           
        left=1mm, right=1mm, top=1mm, bottom=1mm, 
        boxsep=4pt,                        
        before skip=10pt, after skip=10pt, 
        title={\faDatabase~~~\textbf{Prompt: Paper Summarizer}},
        coltitle=white,
        fonttitle=\bfseries\small
    }
}

\tcbset{prompt_parser_filter/.style={
        enhanced,
        breakable,
        colback=prompt,        
        colframe=prompt-frame,            
        fontupper=\normalsize,               
        boxrule=1pt,                       
        arc=4mm,                           
        left=1mm, right=1mm, top=1mm, bottom=1mm, 
        boxsep=4pt,                        
        before skip=10pt, after skip=10pt, 
        title={\faDatabase~~~\textbf{Prompt: Figure Filter}},
        coltitle=white,
        fonttitle=\bfseries\small
    }
}

\tcbset{prompt_planner_matching/.style={
        enhanced,
        breakable,
        colback=prompt,        
        colframe=prompt-frame,            
        fontupper=\normalsize,               
        boxrule=1pt,                       
        arc=4mm,                           
        left=1mm, right=1mm, top=1mm, bottom=1mm, 
        boxsep=4pt,                        
        before skip=10pt, after skip=10pt, 
        title={\faDatabase~~~\textbf{Prompt: Asset Matching}},
        coltitle=white,
        fonttitle=\bfseries\small
    }
}

\tcbset{prompt_planner_painter/.style={
        enhanced,
        breakable,
        colback=prompt,        
        colframe=prompt-frame,            
        fontupper=\normalsize,               
        boxrule=1pt,                       
        arc=4mm,                           
        left=1mm, right=1mm, top=1mm, bottom=1mm, 
        boxsep=4pt,                        
        before skip=10pt, after skip=10pt, 
        title={\faDatabase~~~\textbf{Prompt: Painter}},
        coltitle=white,
        fonttitle=\bfseries\small
    }
}

\tcbset{prompt_planner_commenter/.style={
        enhanced,
        breakable,
        colback=prompt,        
        colframe=prompt-frame,            
        fontupper=\normalsize,               
        boxrule=1pt,                       
        arc=4mm,                           
        left=1mm, right=1mm, top=1mm, bottom=1mm, 
        boxsep=4pt,                        
        before skip=10pt, after skip=10pt, 
        title={\faDatabase~~~\textbf{Prompt: Commenter}},
        coltitle=white,
        fonttitle=\bfseries\small
    }
}

\tcbset{
    vlm_aesthetics_judge/.style={
        enhanced,
        breakable,
        colback=prompt,        
        colframe=prompt-frame,            
        fontupper=\normalsize,               
        boxrule=1pt,                       
        arc=4mm,                           
        left=1mm, right=1mm, top=1mm, bottom=1mm, 
        boxsep=4pt,                        
        before skip=10pt, after skip=10pt, 
        title={\faDatabase~~~\textbf{Prompt: Aesthetics Quality Judge}},
        coltitle=white,
        fonttitle=\bfseries\small
    }
}
\tcbset{
    vlm_element_judge/.style={
        enhanced,
        breakable,
        colback=prompt,        
        colframe=prompt-frame,            
        fontupper=\normalsize,               
        boxrule=1pt,                       
        arc=4mm,                           
        left=1mm, right=1mm, top=1mm, bottom=1mm, 
        boxsep=4pt,                        
        before skip=10pt, after skip=10pt, 
        title={\faDatabase~~~\textbf{Prompt: Element Quality Judge}},
        coltitle=white,
        fonttitle=\bfseries\small
    }
}

\tcbset{
    vlm_layout_judge/.style={
        enhanced,
        breakable,
        colback=prompt,        
        colframe=prompt-frame,            
        fontupper=\normalsize,               
        boxrule=1pt,                       
        arc=4mm,                           
        left=1mm, right=1mm, top=1mm, bottom=1mm, 
        boxsep=4pt,                        
        before skip=10pt, after skip=10pt, 
        title={\faDatabase~~~\textbf{Prompt: Layout Quality Judge}},
        coltitle=white,
        fonttitle=\bfseries\small
    }
}

\newtcolorbox{definitionbox}{
  colback=gray!10,
  colframe=gray!80,
  boxrule=1pt,
  width=\textwidth,
  enlarge left by=0mm,
  enlarge right by=0mm,
  float*=t,  
  boxsep=1pt,
  arc=2.5mm,
  halign=left,
}

\title{Human-Agent Collaborative Paper-to-Page Crafting}

\author{
    {\bf Qianli Ma}$^{1}$\footnotemark[1] \quad
    {\bf Siyu Wang}$^1$\footnotemark[1] \quad
    {\bf Yilin Chen}$^{1}$\footnotemark[1]\quad 
    {\bf Yinhao Tang}$^{2}$\footnotemark[1]\quad 
    {\bf Yixiang Yang}$^{1}$\quad \\ 
    {\bf Chang Guo}$^{1}$ \quad
    {\bf Bingjie Gao}$^{1}$ \quad
    {\bf Zhening Xing}$^{2}$ \quad
    {\bf Yanan Sun}$^{2}$ \quad
    {\bf Zhipeng Zhang}$^{1}$\footnotemark[2] \\
    \textsuperscript{1}AutoLab, SAI, Shanghai Jiao Tong University 
    \textsuperscript{2}Shanghai AI Laboratory \\
    \normalsize
    \texttt{\{mqlqianli,zhipengzhang\}@sjtu.edu.cn} \\
    Project Page$^{\textcolor{darkblue}{1}}$: \href{https://mqleet.github.io/AutoPage_ProjectPage/}{\textcolor{PineGreen}{https://AutoPage.github.io}}
}

\begin{document}
\maketitle

{
\renewcommand{\thefootnote}{\fnsymbol{footnote}}
\footnotetext[1]{Equal Contribution.}
}

{
\renewcommand{\thefootnote}{\fnsymbol{footnote}}
\footnotetext[2]{Corresponding Author.}
}


\begin{abstract}
In the quest for scientific progress, communicating research is as vital as the discovery itself. Yet, researchers are often sidetracked by the manual, repetitive chore of building project webpages to make their dense papers accessible. While automation has tackled static slides and posters, the dynamic, interactive nature of webpages has remained an unaddressed challenge. To bridge this gap, we reframe the problem, arguing that the solution lies not in a single command, but in a collaborative, hierarchical process. We introduce \textbf{AutoPage}, a novel multi-agent system that embodies this philosophy. AutoPage deconstructs paper-to-page creation into a coarse-to-fine pipeline from narrative planning to multimodal content generation and interactive rendering. To combat AI hallucination, dedicated "Checker" agents verify each step against the source paper, while optional human checkpoints ensure the final product aligns perfectly with the author's vision, transforming the system from a mere tool into a powerful collaborative assistant. To rigorously validate our approach, we also construct \textbf{PageBench}, the first benchmark for this new task. Experiments show AutoPage not only generates high-quality, visually appealing pages but does so with remarkable efficiency in under 15 minutes for less than \$0.1. 
Code and dataset will be released at \href{https://mqleet.github.io/AutoPage_ProjectPage/}{Webpage}\footnote{This page is generated by AutoPage. Refresh to see a new, randomly selected version created by AutoPage.}.

\end{abstract}
\section{Introduction}
\label{sec:intro}

Efficient communication is as crucial to scientific advancement as the generation of new knowledge. Although academic papers are the primary medium for research dissemination, their density can hinder accessibility. To address this, researchers often create project pages to distill their work into accessible summaries, highlight key contributions, and showcase demos. However, this is a manual, repetitive process, typically involving the adaptation of existing templates, which consumes valuable research time and also results in inconsistent quality. This motivates our central question that \textit{Can we automate the generation of high-quality project pages directly from academic papers, thereby freeing researchers to focus on core research tasks?}

\begin{figure}[!t]
            \centering
            \includegraphics[width=\linewidth]{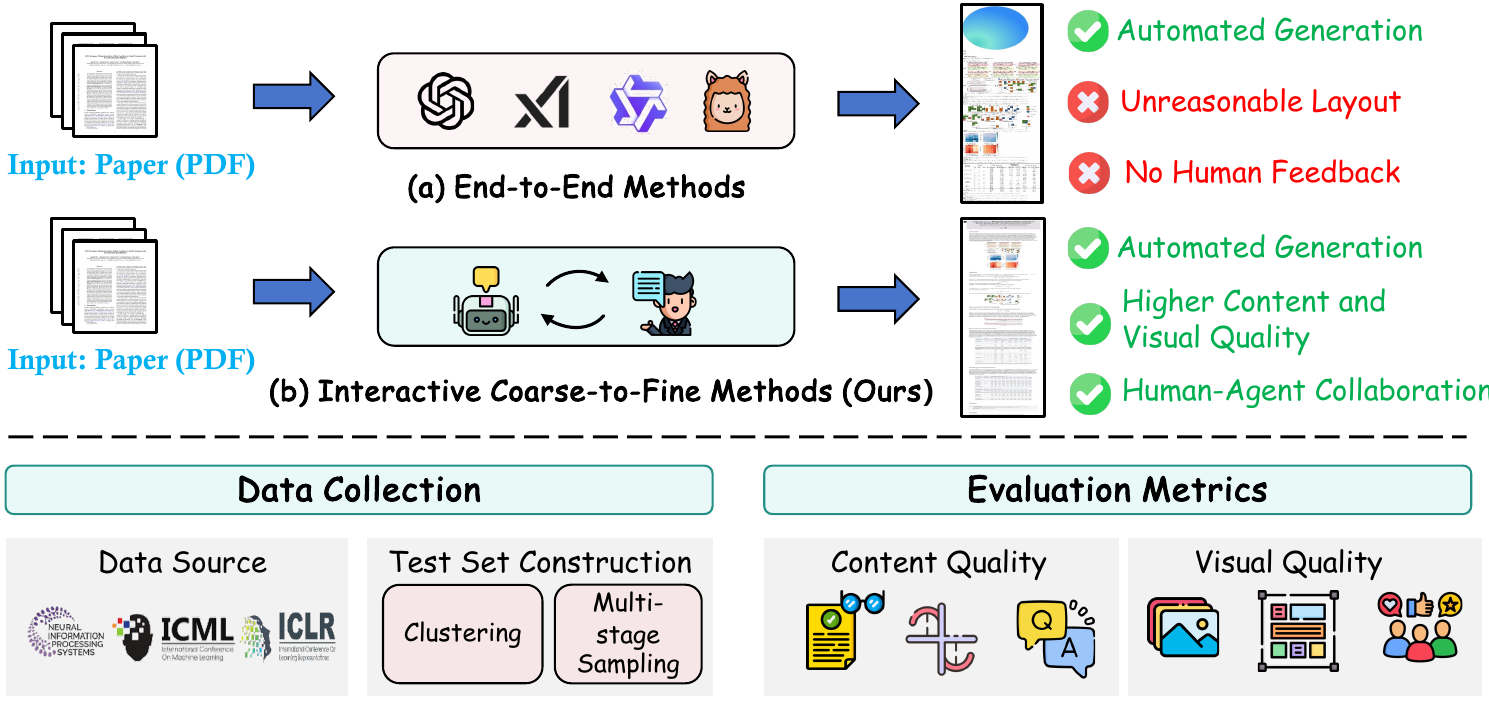}
            \vspace{-20pt}
            \caption{\textbf{Overview of our work.} (a) End-to-end LLMs directly convert papers into project pages, resulting in unreasonable layouts and lacking human feedback. (b) Our proposed AutoPage integrates human-agent collaboration into automated page generation with higher content and visual quality. The figure also illustrates the process for constructing the PageBench benchmark.} 
            

     \label{fig:overview}
    \vspace{-15pt}
\end{figure}

Revisiting the related works, we noted that the automation of research communication has hitherto been confined to static media. More specifically, prior efforts have successfully employed agent systems to convert papers into static visual formats such as slides~\cite{zheng2025pptagent}, posters~\cite{sun2025p2p,pang2025paper2poster}, and videos~\cite{zhu2025paper2video,liu2025preacher}. However, these approaches operate within constrained and fixed-aspect-ratio canvases where the focus remains primarily on visual composition within a bounded space. In contrast, project webpages demand a fundamental shift from canvas-based design to flow-based architecture. Unlike the paginated nature of slides, webpages require a continuous and responsive structure that adapts to varying viewports. Furthermore, they necessitate web-native interactivity, including navigation bars and collapsible modules to facilitate non-linear information exploration. This structural mismatch renders existing static-layout agents ill-suited for the web and highlights the critical need for a system capable of synthesizing fully functional and interactive DOM structures rather than simple visual summaries.


We experimentally found that addressing this gap demands a fundamental shift away from monolithic, end-to-end pipelines, such as an LLM like GPT-4o (see Sec.~\ref{sec:quantitative_study} for more details), shown in Fig.~\ref{fig:overview}\textcolor{darkblue}{a}. Instead, we conceptualize it as a hierarchical, coarse-to-fine generation process augmented by iterative human-agent collaboration, as illustrated in Fig.~\ref{fig:overview}\textcolor{darkblue}{b}. This core principle enables us to manage the task's inherent complexity by first establishing a global narrative structure before progressively refining multimodal details. Crucially, by integrating human feedback at key stages, our approach ensures authorial control and alignment, reframing the system from a simple autonomous generator into a powerful collaborative assistant.

Guided by this philosophy, we introduce \textbf{AutoPage}, a multi-agent system that instantiates our coarse-to-fine, collaborative framework. AutoPage decomposes the complex task of webpage creation into a structured pipeline encompassing three core phases, including narrative planning, multimodal content generation, and interactive page rendering. To ensure factual accuracy and mitigate the risk of LLM hallucination, each phase concludes with a verification step. Here, dedicated LLM/VLM-based "checkers" act like quality inspectors on an assembly line, validating the generated content against the source paper before it proceeds to the next stage. Furthermore, the system is designed for flexible collaboration. While AutoPage can operate fully autonomously from start to finish, it also provides optional checkpoints for human intervention. This allows authors to steer the narrative, adjust visual elements, or make fine-grained edits if they choose, ensuring the final output is not only automated but also perfectly aligned with their vision.

To rigorously evaluate AutoPage and spur future research, we further construct \textbf{PageBench}, the first benchmark dataset tailored for automated paper-to-page generation. PageBench comprises a diverse collection of over 1,500 academic papers paired with their corresponding human-created project pages, along with a proposed comprehensive evaluation protocol that assesses content accuracy, narrative coherence, and visual design.

Our evaluation reveals that AutoPage exhibits model-agnostic adaptability, functioning with various end-to-end models (\textit{e.g.}, GPT-4o, Gemini, and Qwen) without any adjustments to prompts or configurations. It provides a substantial performance uplift in the task of academic page generation. Critically, the system demonstrates exceptional cost-effectiveness and speed, with a single page generated in under 15 minutes for less than \$0.1 using Gemini-2.5-Flash.

Our contributions are fourfold:
$\spadesuit$ We introduce the novel task of automated webpage generation for an academic paper. \ding{170}~We propose an innovative coarse-to-fine, collaborative, and user-friendly framework, implemented as a multi-agent pipeline that integrates LLM/VLM-based ``Checker'' agents for quality control and supports optional human oversight for author alignment. $\clubsuit$~We introduce PageBench, the first benchmark for this task, featuring a suite of novel metrics for holistically evaluating webpages on factual consistency, aesthetics, and authorial alignment. \ding{169}~We demonstrate through extensive evaluations that AutoPage effectively generates factually accurate, visually appealing, and high-quality project pages.

\section{Related Works}


\noindent\textbf{LLM Agents.}~
The story of Large Language Models~(LLMs) is no longer one of solitary genius. They have broken free from their shells as standalone systems, stepping into the role of intelligent ``agents'' within dynamic, collaborative frameworks~\cite{wang2024survey,xi2025rise,xie2024large,zhao2023survey,ge2025autopresent}. This evolution equips them with the autonomy to tackle complex, multi-step tasks once thought to be the exclusive domain of human intellect. Works like ReAct~\cite{yao2023react} have shown how these agents can now autonomously plan strategies~\cite{huang2024understanding,sun2023adaplanner}, wield digital tools~\cite{qu2025tool,shi2025tool}, and reason through intricate problems~\cite{fu2023improving}. At their core, these agents operate in a sophisticated loop. Specifically, they deconstruct abstract goals into actionable plans, enrich their understanding by retrieving external knowledge~\cite{li2025webthinker,li2025search}, and critically improve their own work through self-reflection~\cite{shinn2023reflexion} or even by collaborating with other agents~\cite{tran2025multi}. This powerful paradigm is already reshaping the landscape of scientific discovery. We've seen them act as tireless research assistants, automating literature surveys~\cite{wang2024autosurvey}, aiding in scholarly writing~\cite{weng2025cycleresearcher,ma2026paper2rebuttal}, and ensuring experimental reproducibility~\cite{seo2025paper2code} by translating unstructured material into coherent outputs. Yet, the story doesn't end when the research is complete. A compelling new chapter is now unfolding, focusing on the crucial ``post-research'' phase of communication and dissemination. There is a growing recognition that these same agentic systems, supported by mature development frameworks like LangChain~\cite{Chase_LangChain_2022}, AutoGen~\cite{wu2024autogen}, and Voyager~\cite{wang2023voyager}, can be harnessed to present and share research findings. This emerging trend promises to boost researcher productivity and amplify the impact of their work, moving beyond discovery to ensure it is effectively heard and understood.

\begin{figure*}[!t]
            \centering
            \includegraphics[width=\textwidth,
            trim={0mm 10mm 0mm 0mm}, 
            clip]{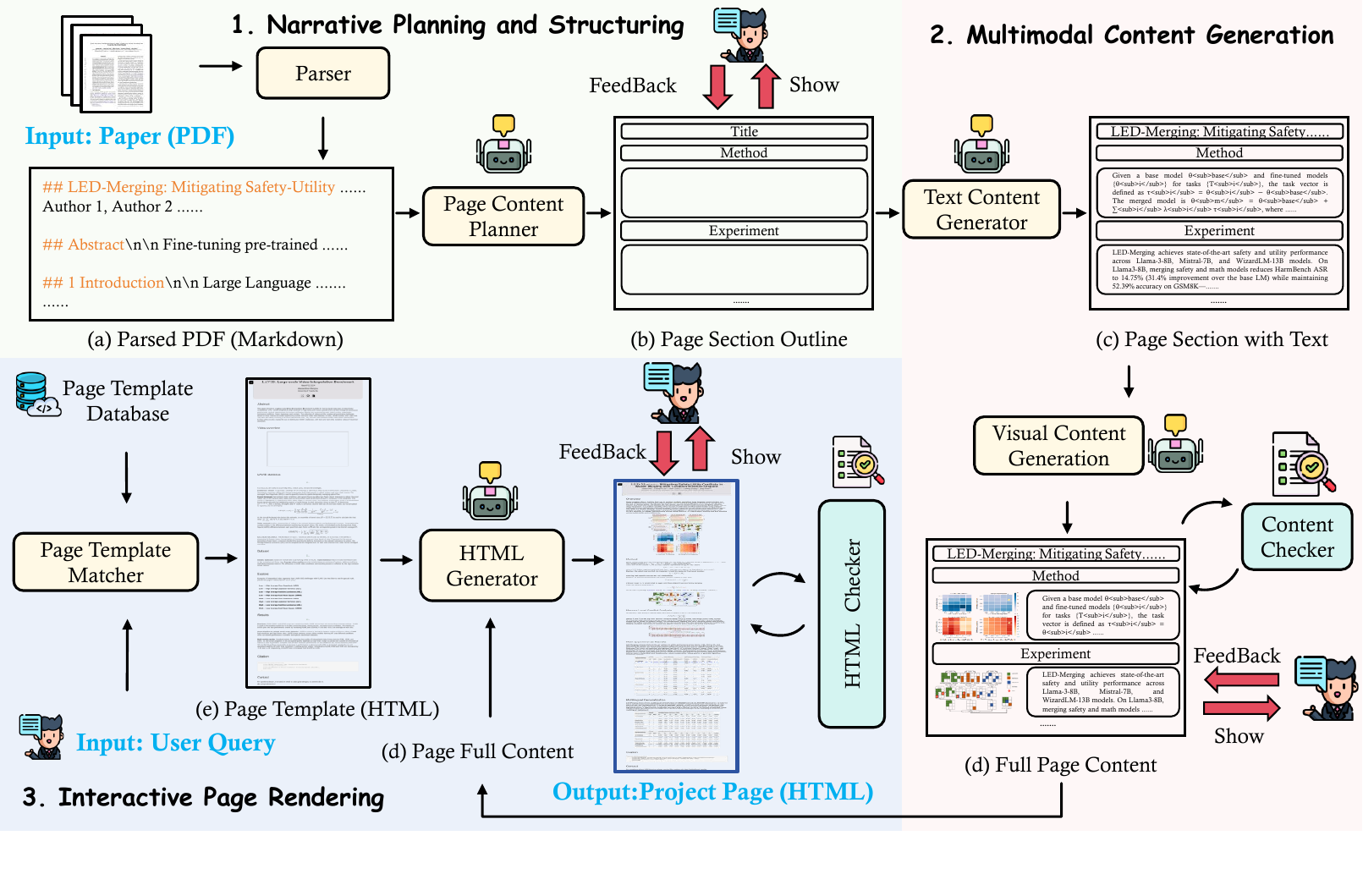}
            \vspace{-20pt}
            \caption{\textbf{Overview of AutoPage.} 
            AutoPage conducts a multi-agent pipeline for transforming papers into interactive webpages:
            (1) \textit{Narrative Planning and Structuring} parses PDFs into Markdown and generates section-level outlines;
            (2) \textit{Multimodal Content Generation} produces coherent text–visual sections;
            (3) \textit{Interactive Page Rendering} matches templates, compiles full HTML pages, and performs final layout checks.
            Throughout all phases, AutoPage integrates verification mechanisms and optional human-in-the-loop checkpoints for reliable and flexible generation.}
     \label{fig:method}
    \vspace{-15pt}
\end{figure*}

\noindent\textbf{Automated Presentation Generation.}~
As aforementioned, a crucial aspect of the productivity enhancement lies in streamlining the creation of visual artifacts for research dissemination. Early, rule-based pipelines~\cite{hu2013ppsgen,paramita2016tailored} represent the first attempt, but their rigid, template-driven nature means they are often brittle and struggle to weave a cohesive narrative from complex scholarly text. A new chapter begins with the rise of agentic systems powered by Vision-Language Models (VLMs), bringing fresh vitality to this field. This evolution is best witnessed in the journey of automated poster generation. Initial efforts like PosterBot~\cite{xu2022posterbot} show promise with neural summarization but are constrained by simple layouts and low-resolution visuals. The true breakthrough comes with sophisticated multi-agent systems like those in Paper2Poster~\cite{pang2025paper2poster} and P2P~\cite{sun2025p2p}. These advanced agents can autonomously plan layouts, write rendering code, and even use VLM feedback to self-correct visual errors. Similar agent-driven innovation has also reshaped slide generation, with tools like PPTAgent~\cite{zheng2025pptagent} and SlideSpawn~\cite{kumar2024slidespawn} demonstrating modular designs that create high-fidelity slides. While slides and posters have received attention, the modern project webpage, arguably today's most vital format for online dissemination, remains surprisingly unexplored territory. To bridge this gap, we propose an agent-driven, automated, and interactive paper-to-page generation system to further enhance researcher productivity.

\section{AutoPage}
\label{sec:pageagent}

This section details the workflow of our \textbf{AutoPage}, including Narrative Planning~(Sec.~\ref{sec:planning_phase}), Multimodal Content Generation~(Sec.~\ref{sec:generation_phase}), and Interactive Page Rendering~(Sec.~\ref{sec:rendering_phase}), as illustrated in Fig.~\ref{fig:method}. Critically, our design incorporates a verification mechanism at the end of each phase to ensure factual grounding, and optional human-in-the-loop checkpoints for flexible collaboration.

\subsection{Narrative Planning and Structuring}
\label{sec:planning_phase}
The initial step in AutoPage is to transform the source paper in PDF format into a structured narrative blueprint for the webpage. This process is orchestrated by two collaborating agents that first deconstruct the paper's content and then architect a new, web-centric narrative.

The process begins with the \textbf{Paper Content Parser}, which ingests the source document and systematically deconstructs it. Using tools like MinerU~\cite{wang2024mineruopensourcesolutionprecise} and Docling~\cite{Docling}, it first converts the document into a raw Markdown format, which is then refined by an LLM into a clean, \texttt{json-like} structure.
The result is an asset library that neatly organizes the paper's core components, which contains: (i) text-based representations, mapping section headings to paragraph-level summaries, and (ii) visual-related representations, linking figures and tables to their corresponding captions and image files.

Building upon this semantically rich asset library, the \textbf{Page Content Planner} then architects the webpage's high-level structure. Rather than performing a simple one-to-one mapping of the paper's original sections, the Planner devises a compelling narrative flow optimized for webpage presentation. It proposes a logical outline, which mirrors how a human would first establish a layout before filling in the details. The output of this stage is a foundational blueprint, which undergoes a verification step to ensure its completeness and logical soundness before proceeding to content generation.

\subsection{Multimodal Content Generation}
\label{sec:generation_phase}
Once the narrative blueprint is finalized, the system proceeds to populate this structure with rich, multimodal content. This task is orchestrated by the \textbf{Page Content Generator}, which operates on a deliberate "text-first" principle. This principle dictates that the narrative prose is generated first, serving as the anchor for the subsequent selection and placement of visual elements, thus ensuring a tight semantic alignment between them.
The process begins with a \textbf{Text Content Generator} sub-component. For each section defined in the blueprint, this component synthesizes the key information from the parser's asset library, transforming it into polished, human-readable paragraphs. Its role is not merely to extract, but to craft a clear and compelling narrative tailored for the web, forming the textual backbone of the page.
With this textual backbone in place, the \textbf{Visual Content Generator} is activated. It analyzes the finalized prose of each section to select and render the most relevant figures or tables from the asset library. This text-driven approach guarantees that each visual element directly supports the accompanying narrative, rather than appearing as a disconnected object, resulting in a coherent module of information.

To ensure the fidelity and quality of the generated content, a \textit{two-stage verification and refinement} process is employed. 
First, an automated \textbf{Content Checker} verifies the consistency between the generated text and its paired visuals. Then, the system offers a crucial checkpoint for \textbf{Human-in-the-Loop Refinement}. At this stage, authors can provide language feedback (\textit{e.g.}, "delete this section", "reorder the sections") to iteratively refine the content until it perfectly aligns with their intent. 
The outcome is a collection of author-approved content modules, ready for final rendering.

\subsection{Interactive Page Rendering}
\label{sec:rendering_phase}
With the author-approved content modules finalized, the system begins the process of rendering them into a polished and interactive webpage.
The process is driven by the \textbf{Page Template Matcher}, which operates on a curated library of templates, each annotated with descriptive tags characterizing its layout and aesthetic properties (\textit{e.g.}, ``background\_color'', ``has\_navigation''). 
Instead of the system making an autonomous choice, the user can specify their stylistic preferences by selecting a combination of these tags.
The agent then filters the library to present only those templates that match the chosen attributes, allowing the user to select the final design.
Once a template is chosen, the system integrates the content modules into its structure and incorporates interactive features. This complete package is then passed to the \textbf{HTML Generator}, which renders the final web artifacts including the HTML, CSS, and JavaScript files.

The process also concludes with a crucial \textit{verification and customization} stage.
First, an automated \textbf{HTML Checker} inspects the rendered page for layout and visual integrity, flagging potential issues like oversized images or color clashes. 
Following this, a final \textbf{Human-in-the-Loop} mechanism is enabled. Authors can provide direct language commands (\textit{e.g.}, "add the navigation bar", "adjust the table colors to match the theme") to fine-tune the webpage styles, ensuring the webpage's visual presentation is polished and precise.

It is worth noting that all the aforementioned human-in-the-loop interactions are optional, as the system can operate in a fully autonomous mode. For instance, a template can be arbitrarily specified or randomly selected by the system without requiring any author intervention. We provide this interactive functionality primarily to enhance user control and flexibility, acknowledging that no agent-based system can be infallible. This optional oversight allows authors to make final corrections and ensure the output perfectly aligns with their vision.

\section{PageBench}

\subsection{Dataset Curation}
\noindent\textbf{Data Source.}~
The dataset of PageBench is curated from project pages associated with papers from three top-tier AI conferences, including NeurIPS, ICML, and ICLR, spanning the years 2023 to 2025. Our curation process involved collecting over 1,500 project pages, followed by a meticulous manual filtering to ensure each entry was a valid project homepage.
The resulting benchmark is a curated corpus rich with multimodal content, including text, figures, and interactive demos, intended to facilitate the development and evaluation of automated project page generation systems.


\noindent\textbf{Test Set and Template Library Construction.}~
To ensure diversity and representativeness, we employed a two-stage sampling strategy to construct a test set and a template library.
First, to build the test set, we extracted structural and stylistic features from our entire corpus, then applied dimensionality reduction and clustering to group similar pages. By sampling from these clusters, we selected 100 pages that represent a broad range of observed page archetypes, forming our primary test set.
Second, to create the template library, we deduplicated this test set using a multi-stage algorithm. 
This process yielded a final, curated Template Library of 87 stylistically distinct pages. The detailed procedure of the sampling strategy for the test set and template library is described in Appendix~\ref{sec:sampling_test_template}.

\subsection{Evaluation Metrics}
To comprehensively evaluate the quality of the generated webpages, we designed a suite of metrics that assess two primary dimensions: \textit{Content Quality} and \textit{Visual Quality}. These metrics collectively measure a model's proficiency in both accurately conveying information and delivering a visually coherent and pleasing presentation.

\subsubsection{Content Quality}
\noindent\textbf{Readability.}~
We evaluate the linguistic fluency and coherence of the generated textual content using an LLM-based readability assessment. Traditional metrics such as perplexity (PPL) primarily capture token-level predictability and have been shown to correlate weakly with human-perceived readability on technical documents. In contrast, LLM-as-a-judge evaluations better reflect discourse-level clarity and coherence. Further details of our evaluation protocol are provided in~\ref{sec:appendix_pagebench_read}.

\noindent\textbf{Semantic Fidelity.}~
To ensure the generated content accurately reflects the source material, we evaluate Semantic Fidelity. This metric measures the semantic correspondence between each webpage section and its original paragraph from the source document. The score is derived by first aligning generated-to-source text pairs and then computing the cosine similarity of their vector embeddings. 
A higher score indicates that the generated content faithfully preserves the meaning of the original text. See Appendix~\ref{sec:appendix_pagebench_semantic} for more details.


\noindent\textbf{Compression-Aware Information Accuracy.}~
Inspired by~\citet{pang2025paper2poster}, we introduce Compression-Aware Information Accuracy to evaluate factual preservation under content compression.
This metric is evaluated using a question-answering (QA) pipeline where we first generate questions from the source paper and then use the generated webpage's text to answer them.
The final score synthesizes two dimensions: the accuracy of the answers (factual correctness) and the text compression ratio (conciseness).
This approach rewards models that produce content that is not only factually accurate but also efficiently summarized. See Appendix~\ref{sec:appendix_pagebench_qa} for more details.

\begin{table*}[hbtp]
\centering
\caption{\textbf{Main evaluation results across our full suite of PageBench.} The best performance among all methods for each metric is in \textbf{bold}, and the second best is \underline{underlined}. For ease of comparison, AutoPage and its corresponding proprietary base models are highlighted in matching colors. AutoPage improves both content and visual quality over different base models, validating its effectiveness in producing accurate, coherent, and visually refined webpages. 
}
\vspace{-10pt}
\label{tab:main_results}
\resizebox{\textwidth}{!}{
\begin{tabular}{lcc|ccc|ccc}
\toprule
\multirow{2}{*}{\textbf{Method}} & \multirow{2}{*}{\textbf{System Type}} & \multirow{2}{*}{\textbf{Open-Source}} & \multicolumn{3}{c|}{\textbf{Content Quality}} & \multicolumn{3}{c}{\textbf{Visual Quality}} \\
\cmidrule(lr){4-6} \cmidrule(lr){7-9}
 &  &  & \begin{tabular}[c]{@{}c@{}}Readability \textbf{$\uparrow$}\end{tabular} & \begin{tabular}[c]{@{}c@{}}Semantic\\ Fidelity \textbf{$\uparrow$}\end{tabular} & \begin{tabular}[c]{@{}c@{}}Comp.-Aware \\ Info. Acc. \textbf{$\uparrow$} \end{tabular} & \begin{tabular}[c]{@{}c@{}}Visual Content\\ Accuracy \textbf{$\uparrow$} \end{tabular} & \begin{tabular}[c]{@{}c@{}}Layout and\\ Cohesion\textbf{$\uparrow$}\end{tabular} & \begin{tabular}[c]{@{}c@{}}Aesthetic\\ Score\textbf{$\uparrow$}\end{tabular} \\
\midrule
\quad GPT-OSS-120B & E2E &\textcolor{gray}{\usym{2714}} & 2.96 & 0.608 & 1.719 & 2.74 & 1.82 & 2.65 \\
\quad llama-3.1-70B &E2E  &\textcolor{gray}{\usym{2714}} & 2.58 & 0.442 & 1.270 & 2.52 & 1.78 &2.41 \\

\quad Grok4-fast &E2E  & & 3.06 & 0.603 & 1.808 & 2.68 & 2.01 & 2.67 \\
\quad GLM-4.5-Air & E2E & &3.04 & 0.587 & 1.788 & 2.98 & 2.03 & 2.67 \\

\midrule
\quad Qwen3-235B-A22B &E2E  &\textcolor{gray}{\usym{2714}} & 3.0 & 0.571 & \underline{1.890} & 2.52 & 1.93 & 2.46 \\
\rowcolor{lemonchiffon}\quad AutoPage-Qwen &Multi-Agent & & 3.14 & 0.663 & 1.837 & 3.01 & 2.28 & 2.72 \\
\midrule
\quad GPT4o-mini &E2E  & & 2.85 & 0.554 & 1.786 & 2.96 & 2.08 & 2.71 \\
\rowcolor{lightcoral}\quad AutoPage-GPT4o-mini &Multi-Agent  & & \underline{3.33} & 0.621 & \textbf{1.941} & 3.08 & \underline{2.38} & \underline{2.95} \\
\midrule
\quad Gemini2.5-Flash &E2E  & & 3.0 &0.684 & 1.276 & 2.82 & 2.00 & 2.48 \\
\quad AutoGen-Gemini2.5-Flash &Multi-Agent  & & 3.10 &\underline{0.711}& 1.330 & 2.61 & 1.80 & 2.54 \\
\rowcolor{skyblue}\quad AutoPage-Gemini2.5-Flash & Multi-Agent & & 3.17 & 
\textbf{0.742} & 1.591 & \textbf{3.13} & 2.15 & 2.69 \\
\midrule
\quad GPT5-mini & E2E & &  3.15 & 0.631 & 1.589 & 2.99 &2.12 & 2.74 \\
\rowcolor{SeaGreen} \quad AutoPage-GPT5-mini &Multi-Agent & &\textbf{3.38} & 0.702 & 1.818 & \underline{3.11} &\textbf{2.40} & \textbf{2.96} \\

\bottomrule
\end{tabular}
}
\vspace{-15pt}
\end{table*}

\subsubsection{Visual Quality}
To evaluate the overall visual quality of the generated pages, we employ a unified VLM-as-Judge framework.
For each dimension, the VLM is prompted to act as a specialized, strict reviewer, assigning a score on a five-point scale. The detailed prompts and scoring rubrics used for these evaluations are provided in Appendix~\ref{sec:appendix-prompt-templates}.

\noindent\textbf{Visual Content Accuracy.}~
This metric assesses the correct presentation of critical visual elements on the page. The evaluation focuses strictly on two objective criteria: the correct rendering of mathematical formulas and the contextual relevance of images to their surrounding text.


\noindent\textbf{Layout and Cohesion.}~
This metric evaluates the structural integrity and visual balance of the page.
The assessment focuses on identifying flaws that disrupt the visual flow and professional appearance, such as disproportionately scaled images and the presence of excessive or poorly distributed white space, aiming to reward designs that offer a coherent and comfortable reading experience.

\noindent\textbf{Aesthetic Score.}~
This dimension quantifies the page's holistic visual appeal by assessing its overall aesthetic feel.
The evaluation focuses on design harmony, including the coherence of the color scheme, style consistency, and clarity of the visual hierarchy. Distinct from content-focused metrics, this score reflects the page's overall visual impression and professional polish.

\section{Experiments}
\label{sec:experiments}

\subsection{Experiment Setup}
We assess the efficacy of the proposed AutoPage by benchmarking it against a diverse set of advanced LLMs, and then ablating the influence of each component. It is important to note that for the purpose of scalable evaluation, all subsequent experiments were conducted using AutoPage in an exclusively automated fashion, without human intervention. The human-in-the-loop configuration, while potentially beneficial, is infeasible for batch testing. Therefore, the performance metrics reported here should be interpreted as a \textit{conservative lower bound} of our system's full potential. We will supplement the App.~\ref{sec:appendix-human-in-the-loop} with experimental evidence demonstrating the effectiveness of human-in-the-loop.


\noindent\textbf{Baselines.}~
The compared baseline models can be categorized into:
\textit{Closed-Source Models} including GPT-4o-mini~\cite{achiam2023gpt}, Grok-4-fast~\cite{xGrok},  GLM-4-Air~\cite{glm2024chatglm} and GPT-5-mini~\cite{openai2025gpt5}.
\textit{Open-Source Models} including Qwen3-235B~\cite{qwen3}, GPT-OSS-120B~\cite{openai2025gptoss120bgptoss20bmodel}, and Llama-3.1-70B~\cite{llama3_1}. 
\textit{Multi-Agent System} including AutoGen~\cite{wu2024autogen}, which is a powerful multi-agent baseline.


\noindent\textbf{Avoiding Information Leakage in Evaluation.}
To accurately evaluate each model's ability to automatically generate project webpages, we designed an evaluation protocol to prevent a key form of information leakage. The potential issue is that a model might copy content directly from the provided webpage template, rather than synthesizing it from the source paper's content. Such behavior would not reflect true generative understanding. Therefore, to ensure a fair assessment, we decouple the paper's content from its original webpage layout. In our setup, each model must generate a webpage for a given paper using a template derived from a completely different paper's project website. This cross-pairing strategy forces the model to rely on the source document to generate content, providing a more rigorous test of its capabilities. We further examine a template-free variant of this setting in Appendix~\ref{sec:appendix_template_free}, where all systems generate webpages without receiving any preset layout.

\subsection{Main Results}
\label{sec:quantitative_study}

\noindent\textbf{AutoPage enhances end-to-end methods.}~
A key finding from our experiments is that AutoPage acts as a powerful enhancer for existing end-to-end methods, significantly elevating both their content and visual generation quality. As detailed in Tab.~\ref{tab:main_results}, we observed consistent and substantial performance gains compared to their respective end-to-end baselines.
For instance, when paired with GPT4o-mini, AutoPage-GPT4o-mini surpasses the baseline GPT4o-mini across all evaluated metrics. Notably, it boosts the Aesthetic Score from 2.71 to 2.95 and improves Layout and Cohesion from 2.08 to 2.38, demonstrating its superior capability in generating visually appealing and well-structured webpages.
A similar trend is observed with AutoPage-Gemini-2.5-Flash, which achieves a higher Semantic Fidelity~(0.742 \textit{v.s.} 0.684) and Visual Content Accuracy (3.13 \textit{v.s.} 2.82) compared to the baseline Gemini-2.5-Flash. 
Furthermore, it shows a dramatic improvement in Compression-Aware Information Accuracy, jumping from 1.276 to 1.941, which is the highest score achieved for this metric across all tested methods.
This pattern also holds for the open-source model, where AutoPage-Qwen shows marked improvements over the end-to-end Qwen baseline, particularly in all three Visual Quality metrics. 
These results collectively validate that AutoPage is an effective and versatile framework that can augment various large models, systematically enhancing their ability to produce higher-quality webpages.

\noindent\textbf{AutoPage Narrows the Performance Gap Across Backbones.}~
An interesting finding is AutoPage's ability to narrow the performance gap between backbone models of varying capabilities. 
As shown in Tab.~\ref{tab:main_results}, a significant performance gap initially exists between the stronger Gemini-2.5-Flash and the weaker open-source Qwen, \textit{e.g.,} in Visual Content Accuracy (2.82 \textit{v.s.} 2.52).
However, AutoPage acts as a great equalizer. While it enhances both models, its impact is far more transformative for the weaker backbone. Specifically, AutoPage boosts Qwen's Visual Content Accuracy score by 0.49 (from 2.52 to 3.01), a gain substantially larger than the 0.31 seen for Gemini-2.5-Flash. This disproportionate improvement slashes the initial performance gap by more than half (0.30 to 0.12). 
This demonstrates that AutoPage is not just an incremental add-on for strong models, but a transformative component that elevates weaker backbones to a competitive state.

\begin{figure}[!t]
            \centering
            \includegraphics[width=\linewidth]{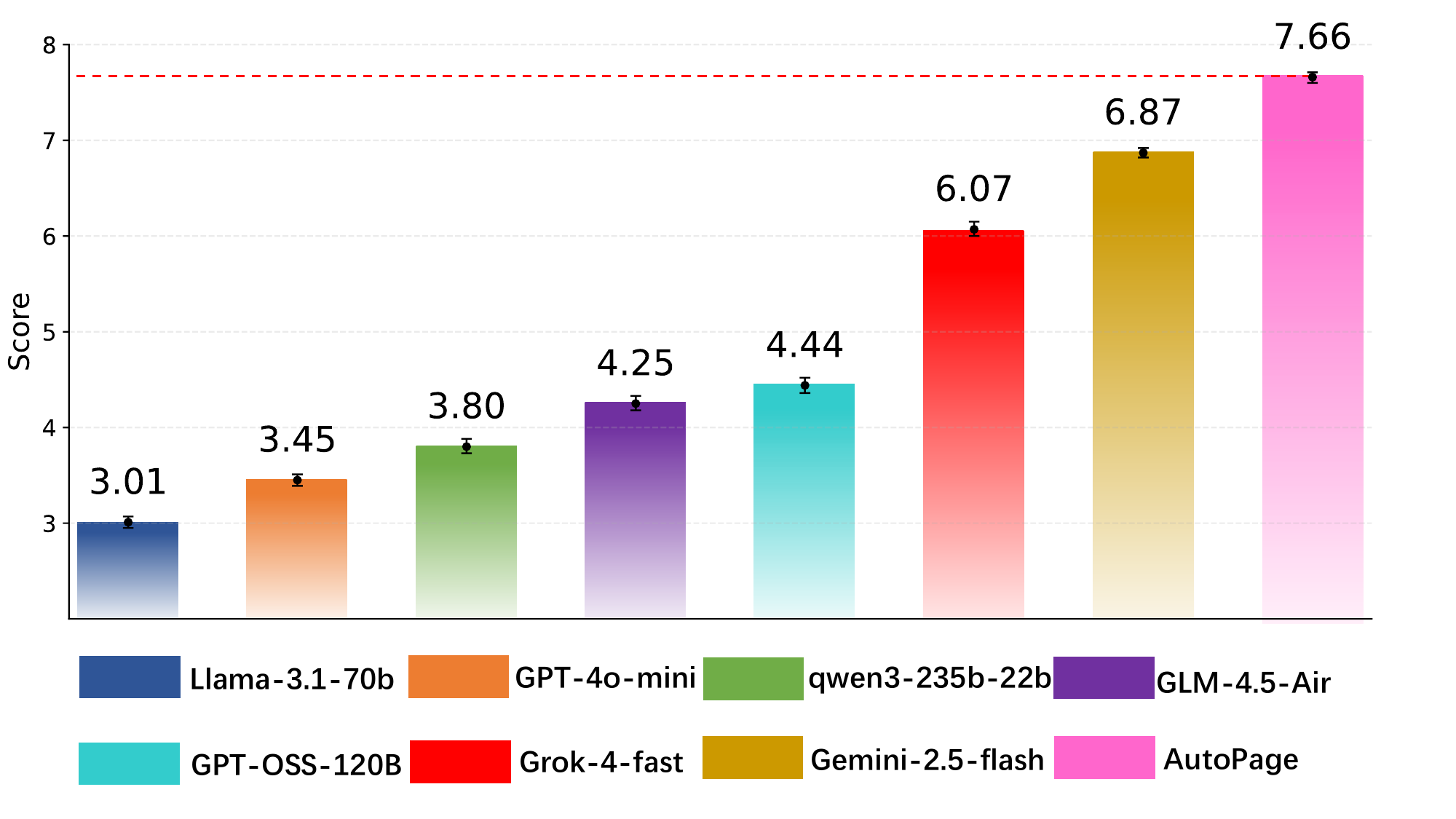}
            \vspace{-20pt}
            \caption{\textbf{Human preference study.} The bar chart shows that AutoPage attains the highest user preference score, surpassing all baselines with more informative, coherent, and visually engaging webpages. Error bars indicate 95\% confidence intervals.} 
            
     \label{fig:user-study}
     \vspace{-15pt}
\end{figure}

\noindent\textbf{Beyond Content: AutoPage as a Visual Architect.}~
Beyond ensuring high content fidelity, AutoPage demonstrates exceptional strength as a "visual architect," skillfully enhancing the aesthetic and structural quality of the generated pages. This layered capability is evident on the Qwen model. The framework secures a respectable 16\% gain in Semantic Fidelity (from 0.571 to 0.663), and in addition to this, delivers a powerful leap in visual metrics. Specifically, Visual Content Accuracy soars by nearly 20\% (from 2.52 to 3.01).
This mastery of the visual dimension is not just a quantitative improvement but translates directly into a superior user experience. This conclusion is decisively supported by our user study (Sec.~\ref{sec:user_study}), as shown in Fig.~\ref{fig:user-study}, our method achieved the highest average score (7.66 out of 10), confirming that its superior visual architecture results in a product that is tangibly more appealing and usable to humans.

\begin{figure*}[!t]
            \vspace{-20pt}
            \centering
            \includegraphics[width=\textwidth]{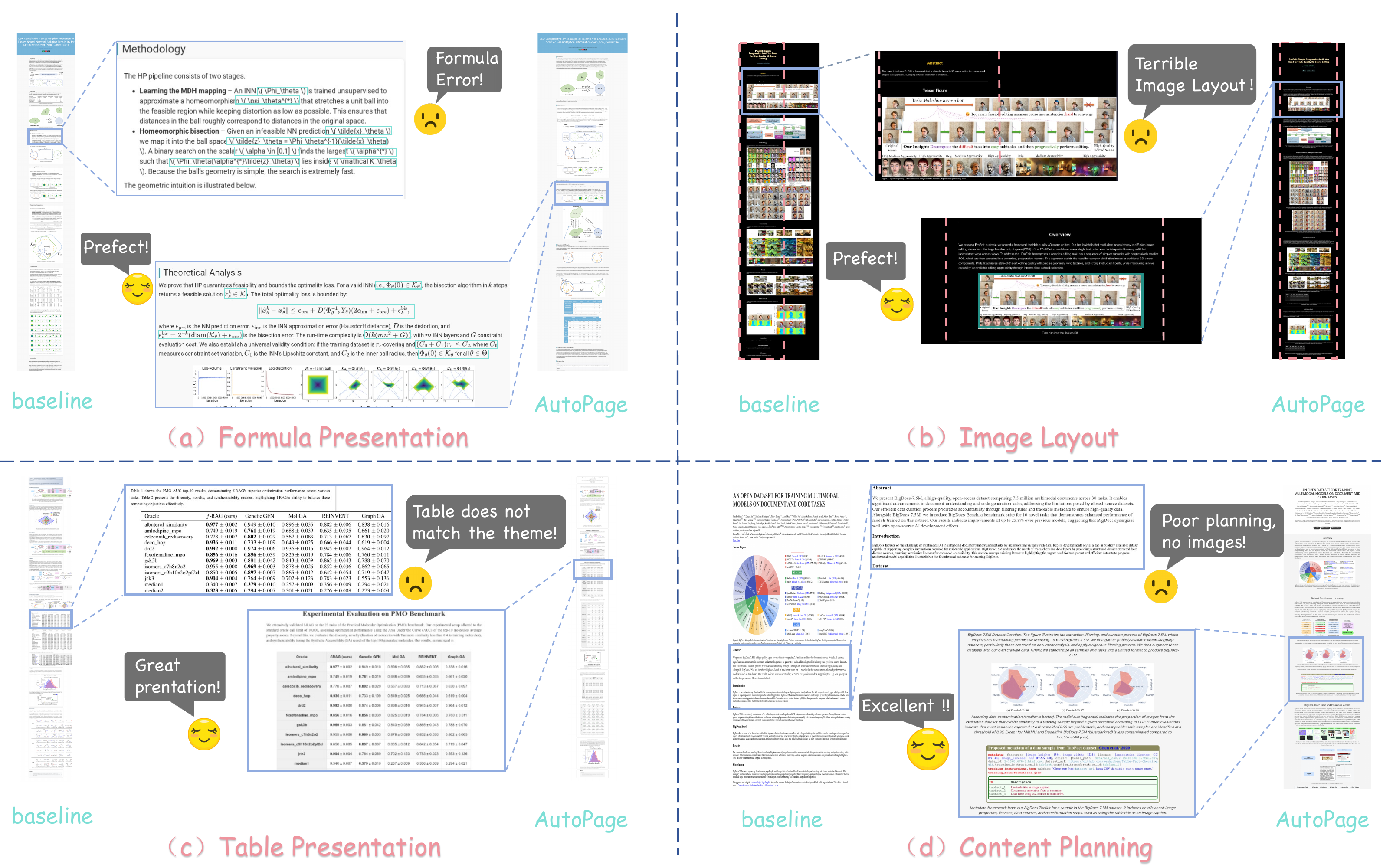}
            \vspace{-10pt}
            \caption{\textbf{Qualitative comparison illustrating AutoPage’s superior generation quality over baselines.} The figure highlights four common scenarios where AutoPage demonstrates superior performance: (a) Formula Presentation; (b) Image Layout; (c) Table Presentation; (d) Content Planning. This qualitative comparison demonstrates AutoPage's ability not just to fill a page with content, but to thoughtfully design it.}
     \label{fig:visual_results}
    \vspace{-15pt}
\end{figure*}

\subsection{User Study}
\label{sec:user_study}
To evaluate the human-perceived quality of the generated webpages, we conducted a user study with 20 participants.
For each source paper, participants were shown a group of 8 webpages generated by the different models.
A key aspect of our methodology was a \textit{forced-choice rating mechanism}. To elicit fine-grained distinctions, participants were required to assign a unique score from 1 (Completely Unusable) to 10 (Perfect) to each of the 8 pages within a group. 
This constraint effectively compelled them to create a relative ranking, preventing score clustering and providing a clearer signal of preference. Further details on the study protocol , the complete scoring rubric and why we adopt force-choice rating mechanism are available in Appendix~\ref{sec:appendix-user-study}.
Our user study demonstrates that webpages generated by AutoPage are preferred by human evaluators.
As shown in Fig.~\ref{fig:user-study}, AutoPage achieved the highest average score of 7.66, establishing a clear performance hierarchy.
It not only leads but also maintains a discernible advantage over other strong models like \texttt{Grok4-fast} (6.07) and \texttt{Gemini2.5-Flash} (6.87). Furthermore, the substantial gap between our model and the lower-scoring models(\textit{e.g.} \texttt{GPT4o-mini}~(3.45)) underscores the difficulty of the task and validates AutoPage's superior ability to produce webpages that better align with human expectations.
Notably, this study did not incorporate human-in-the-loop feedback. We believe introducing this could further improve PageAgent's scores by more directly aligning the generated webpages with human preferences.

\subsection{Case Study}
\label{sec:qualitive study}
As shown in Fig.~\ref{fig:visual_results}, we present several cases that highlight the qualitative leap AutoPage provides over standard end-to-end methods. 
For instance, baseline end-to-end methods frequently fail to correctly render mathematical formulas and resort to inefficient, vertically stacked layouts for image galleries, as shown in Fig.~\ref{fig:visual_results}\textcolor{darkblue}{a} and Fig.~\ref{fig:visual_results}\textcolor{darkblue}{b}.
Beyond ensuring the correct presentation of intricate elements like formulas, AutoPage also applies a keen design sense. It organizes images into structured galleries and styles components like tables to match the page's theme, shown in Fig.~\ref{fig:visual_results}\textcolor{darkblue}{c}, avoiding the discordant look produced by the baseline.
Furthermore, AutoPage demonstrates a crucial strategic capability absent in baselines: content planning, shown in Fig.~\ref{fig:visual_results}\textcolor{darkblue}{d}. It analyzes the document's substance and enriches it by generating relevant visualizations, transforming dense text into an engaging and easy-to-digest format.

Beyond the main results, we provide extensive supplementary analyses in the Appendix to offer additional insights into the behavior and design choices of AutoPage. These include systematic ablations on verifier components, template-free generation settings to rule out layout priors, detailed justifications of our evaluation metrics, controlled studies on the effectiveness of human-in-the-loop feedback, and additional visualization result comparisons. Together, these results offer a comprehensive validation of AutoPage’s design choices, robustness, and practical applicability beyond the main experimental findings.


\begin{table*}[t]
\centering
\caption{Ablation study on agent roles. We keep the evaluation dimensions consistent with the main table, and additionally report Compression Rate as an efficiency-oriented metric. Higher is better for all metrics.}
\label{tab:agent_ablation}
\scriptsize
\setlength{\tabcolsep}{4pt}
\renewcommand{\arraystretch}{1.1}
\resizebox{\textwidth}{!}{%
\begin{tabular}{lccc!{\vrule width 0.6pt}ccc!{\vrule width 0.6pt}c}
\toprule
\multirow{2}{*}{\textbf{Method}} 
& \multicolumn{3}{c!{\vrule width 0.6pt}}{\textbf{Content Quality}} 
& \multicolumn{3}{c!{\vrule width 0.6pt}}{\textbf{Visual Quality}} 
& \multirow{2}{*}{\makecell[c]{\textbf{Compression}\\\textbf{Rate}$\uparrow$}} \\
\cmidrule(lr){2-4} \cmidrule(lr){5-7}
& \textbf{Readability}$\uparrow$ 
& \textbf{Semantic Fidelity}$\uparrow$ 
& \makecell[c]{\textbf{Comp.-Aware}\\\textbf{Info. Acc.}$\uparrow$} 
& \makecell[c]{\textbf{Visual Content}\\\textbf{Accuracy}$\uparrow$} 
& \makecell[c]{\textbf{Layout and}\\\textbf{Cohesion}$\uparrow$} 
& \textbf{Aesthetic Score}$\uparrow$ 
&  \\
\midrule
\textbf{AutoPage} 
& 3.18 & 0.739 & \textbf{1.744} & \textbf{3.13} & \textbf{2.15} & \textbf{2.69} & \textbf{9.168} \\
w/o Page Content Planner 
& 3.03 & 0.702 & 1.523 & 2.80 & 1.80 & 2.63 & 8.753 \\
w/o Text Content Generator 
& \textbf{3.26} & \textbf{0.748} & 1.449 & 2.84 & 1.73 & 2.52 & 4.326 \\
\bottomrule
\end{tabular}%
}
\end{table*}
\subsection{Ablation Study}
\label{sec:Ablation_Study}
\paragraph{Effectiveness of Different Agents.}
We analyze the necessity of our three-stage role decomposition by ablating the \textit{Page Content Planner} and \textit{Text Content Generator} in Table~\ref{tab:agent_ablation}. 

Removing the \textit{Page Content Planner} degrades all metrics, especially \textit{Layout \& Cohesion} (2.15 to 1.80) and \textit{Compression-Aware Information Accuracy} (1.744 to 1.523), highlighting the role of global structural planning in webpage organization.

Removing the \textit{Text Content Generator} slightly increases \textit{Readability} and \textit{Semantic Fidelity}, but this is mainly because the system copies more source text verbatim, preserving surface similarity at the cost of heavy redundancy. As a result, the \textit{Compression Rate} drops sharply (from 9.168 to 4.326), while \textit{Compression-Aware Information Accuracy} also decreases to 1.449. This indicates that retaining more raw text does not improve effective information preservation, but instead harms concise and usable webpage generation. 
These results highlight that academic webpage generation depends not only on fluent generation, but also on \emph{compression fidelity}: compressing long documents while preserving answerable evidence and coherent structure. This also supports the design motivation of our compression-aware evaluation metric.

\paragraph{Effectiveness of Checker.}

We evaluate the performance of AutoPage's self-correction mechanism by conducting an ablation study on its verifiers: the full content checker and the HTML checker. The full content checker automatically verifies the consistency and relevance between the generated text and its accompanying visuals. Subsequently, the HTML checker inspects the final page for layout and visual integrity, flagging issues like oversized images or tables and colors that clash with the template's theme.
In our ablation experiment, we systematically disable these components to create three distinct variants: (i) AutoPage \texttt{w/o} full content checker, (ii) AutoPage \texttt{w/o} HTML checker, and (iii) AutoPage \texttt{w/o} all checkers.


\begin{table}[h!] 
\centering
\caption{Ablation study on different verifiers in AutoPage. The Full Content Checker verifies text--visual consistency, while the HTML Checker verifies layout and rendering quality. Removing either verifier degrades performance, and removing both causes the largest drop.}
\vspace{-10pt}
\label{tab:ablation_verfier}
\resizebox{\linewidth}{!}{
\begin{tabular}{l cccc}
\toprule
& & \multicolumn{3}{c}{\textbf{\texttt{w/o} Verifier}} \\
\cmidrule(lr){3-5}
\textbf{Metric} & {\textbf{AutoPage}} & {Full Content} & {HTML} & {All} \\
\midrule
\multicolumn{5}{l}{\textit{Content Quality}} \\
\quad Readability$\uparrow$ & \textbf{3.18} & \underline{3.08} & 3.06 & 3.01 \\
\quad Semantic Fidelity$\uparrow$  & \textbf{0.739} & 0.695 & \underline{}{0.708} & 0.695 \\
\quad Comp.-Aware Info. Acc. $\uparrow$ & \textbf{1.744} & 1.533 & \underline{1.583} & 1.533 \\
\midrule
\multicolumn{5}{l}{\textit{Visual Quality}} \\
\quad Visual Content Acc. $\uparrow$ & \textbf{3.13} & \underline{}{3.05} & 2.90 & 2.75 \\
\quad Layout and Cohesion $\uparrow$ & \textbf{2.15} & \underline{1.95} & 1.65 & 1.60 \\
\quad Aesthetic Score $\uparrow$ & \textbf{2.69} & \underline{2.25} & 2.20 & 1.90 \\
\bottomrule
\end{tabular}
}
\vspace{-15pt}
\end{table}



The results presented in Tab.~\ref{tab:ablation_verfier} show a notable performance degradation across all ablation settings, with the most significant drop observed when both verifiers are removed. For instance, when both verifiers are removed, the Visual Content Accuracy drops from 3.13 to 2.75, and the Aesthetic Score plummets from 2.69 to 1.90. 
Disabling only the HTML checker causes the Layout and Cohesion score to fall sharply from 2.15 to 1.65. Meanwhile, removing the full content checker leads to a degradation in Semantic Fidelity from 0.739 to 0.695.
This result confirms the indispensable value of each of our verifiers and establishes that the entire verifier-driven, multi-turn refinement process is the key to producing high-quality webpages.

\section{Conclusion}
We presented AutoPage, a multi-agent framework that transforms academic papers into interactive project webpages. Together with PageBench, the first benchmark for this task, we enable principled evaluation across fidelity, compression, and aesthetics. Experiments show that AutoPage produces coherent, dynamic, and accessible webpages, lowering the cost of scientific communication. We hope this work provides a foundation for future systems that further expand the accessibility and impact of research.


\section*{Limitations and Future Work}
While AutoPage demonstrates promising results in automatically generating project webpages, there still remain several limitations that point to directions for future exploration.
First, PageBench currently focuses on project pages linked to machine learning conferences (ICML, ICLR, NeurIPS, 2023-2025). While this provides a strong starting point, extending coverage to other venues such as ACL, CVPR, and KDD would enhance diversity and domain generality.
Second, our prototype system relies on commercial API-based models, which may incur monetary costs and limit reproducibility. Future work could incorporate open-source or locally deployable models to reduce cost and improve accessibility.
Finally, while our evaluation captures core aspects such as content fidelity and visual aesthetics, additional user-centric studies could provide a complementary perspective on usability and long-term impact.

\section*{Broader Impact and Ethics Statement}
This work builds on project webpages publicly released by authors of papers at ICML, ICLR, and NeurIPS (2023--2025). All data in PageBench is collected from open sources and is intended solely for academic research purposes. We will release the dataset and benchmark tools openly to facilitate transparency, reproducibility, and further study by the community.
We gratefully acknowledge the authors of the original project pages, whose efforts not only enhance the accessibility of their own work but also enable research such as ours.
By automating webpage generation, our goal is to reduce the cost of scientific communication while preserving author agency over final outputs. 
We believe that sharing both methods and data will contribute positively to openness and inclusivity in the research community.

\section*{Acknowledgments}
The work was supported by the Natural Science Foundation of China (Grant No. 62503323)

\bibliography{custom}

\appendix
\newpage
\appendix
\section{Ablation on Verifiers}
\label{sec:appendix-ablation}


\subsection{Template-Free Generation}
\label{sec:appendix_template_free}
To ensure a fair comparison in the main experiment, both AutoPage and the end-to-end LLM baselines were given the same parsed paper content as well as the same human-authored HTML templates. All baselines were explicitly instructed to generate webpages that conform to these templates, ensuring that AutoPage \textbf{does not benefit from any privileged access to layout priors}.

Human-authored templates are introduced because end-to-end LLMs generating webpages from scratch tend to collapse into highly uniform layouts and model-specific stylistic biases. A diverse template library enriches stylistic variation for all systems and prevents unfair penalization of baselines due to layout collapse.
To further disentangle the contribution of templates from the contribution of our agent pipeline, we conducted an \textit{additional template-free setting}, in which neither system received any preset layout. As shown in Tab.~\ref{tab:template_effect}, AutoPage continues to outperform end-to-end LLMs in this setting, indicating that the gains arise from the structured agent pipeline itself rather than dependence on template retrieval.

\begin{table}[h!]
\centering
\caption{\textbf{Effect of Template Usage on Webpage Generation Quality.} Comparison between end-to-end LLM baselines and AutoPage under both template-free and template-based settings. AutoPage consistently outperforms its corresponding end-to-end LLM across all metrics in both settings. Notably, the performance gains persist even without templates, indicating that the improvements stem from the structured agent pipeline rather than reliance on template guidance.}
\vspace{-6pt}
\label{tab:template_effect}

\renewcommand{\arraystretch}{1.30} 

\resizebox{\linewidth}{!}{
{\LARGE
\begin{tabular}{l l c c c}
\toprule
\textbf{Setting} &
\textbf{Model} &
\makecell{\textbf{Visual Content}\\ \textbf{Accuracy $\uparrow$}} &
\makecell{\textbf{Layout and}\\ \textbf{Cohesion $\uparrow$}} &
\makecell{\textbf{Aesthetic}\\ \textbf{Score $\uparrow$}} \\
\midrule
\multicolumn{5}{l}{\textit{Template-Free}} \\
& gpt-5-mini                & 2.84 & 1.95 & 2.48 \\
& AutoPage-gpt-5-mini       & \textbf{2.86} & \textbf{2.00} & \textbf{2.81} \\
& Gemini-2.5-Flash          & 2.75 & 1.95 & 2.42 \\
& AutoPage-Gemini-2.5-Flash & \textbf{2.94} & \textbf{1.99} & \textbf{2.44} \\
\midrule
\multicolumn{5}{l}{\textit{Template-Based}} \\
& gpt-5-mini                & 2.99 & 2.12 & 2.74 \\
& AutoPage-gpt-5-mini       & \textbf{3.11} & \textbf{2.40} & \textbf{2.96} \\
& Gemini-2.5-Flash          & 2.82 & 2.00 & 2.48 \\
& AutoPage-Gemini-2.5-Flash & \textbf{3.13} & \textbf{2.15} & \textbf{2.69} \\
\bottomrule
\end{tabular}
}}
\vspace{-10pt}
\end{table}

\subsection{Generation in Different Evaluation Scope}

\begin{table*}[t]
\centering
\caption{Generation under different evaluation scopes. We evaluate AutoPage and the end-to-end Gemini-2.5-Flash baseline on two additional settings beyond the original benchmark: non-paper technical reports and non-ML scientific papers. Higher is better for all metrics.}
\label{tab:different_scope}
\scriptsize
\setlength{\tabcolsep}{4pt}
\renewcommand{\arraystretch}{1.08}
\resizebox{\textwidth}{!}{%
\begin{tabular}{lccc!{\vrule width 0.6pt}ccc}
\toprule
\multirow{2}{*}{\textbf{Setting, Model}}
& \multicolumn{3}{c!{\vrule width 0.6pt}}{\textbf{Content Quality}}
& \multicolumn{3}{c}{\textbf{Visual Quality}} \\
\cmidrule(lr){2-4} \cmidrule(l){5-7}
& \textbf{Readability}$\uparrow$
& \textbf{Semantic Fidelity}$\uparrow$
& \makecell[c]{\textbf{Comp.-Aware}\\\textbf{Info. Acc.}$\uparrow$}
& \makecell[c]{\textbf{Visual Content}\\\textbf{Accuracy}$\uparrow$}
& \makecell[c]{\textbf{Layout and}\\\textbf{Cohesion}$\uparrow$}
& \textbf{Aesthetic Score}$\uparrow$ \\
\midrule

\textit{Non-paper Technical Reports} & & & & & & \\
\quad Gemini-2.5-Flash
& 3.04 & 0.668 & 1.327 & 2.82 & 2.30 & 2.20 \\
\quad AutoPage-Gemini-2.5-Flash
& \textbf{3.12} & \textbf{0.728} & \textbf{1.511} & \textbf{2.91} & \textbf{2.35} & \textbf{2.31} \\
\midrule

\textit{Non-ML Scientific Papers} & & & & & & \\
\quad Gemini-2.5-Flash
& 3.02 & 0.696 & 1.339 & 2.77 & 2.06 & 2.53 \\
\quad AutoPage-Gemini-2.5-Flash
& \textbf{3.14} & \textbf{0.732} & \textbf{1.611} & \textbf{2.93} & \textbf{2.12} & \textbf{2.66} \\
\bottomrule
\end{tabular}%
}
\end{table*}
Although the original PageBench mainly focuses on ML papers, AutoPage is intended as a general document-to-webpage generation framework rather than an ML-specific system. To evaluate its robustness under broader evaluation scopes, we further test AutoPage-Gemini-2.5-Flash in two additional settings beyond the original benchmark: (1) 10 non-paper technical reports with documentation-style writing and heterogeneous layouts, and (2) 10 non-ML scientific papers spanning diverse disciplines, including Medicine, Biology, Physics, Chemistry, and Systems. In both settings, we compare against the same end-to-end Gemini-2.5-Flash baseline under identical generation conditions.

As shown in Table~\ref{tab:different_scope}, AutoPage consistently outperforms the baseline across all metrics in both settings. On non-paper technical reports, AutoPage improves Semantic Fidelity from 0.668 to 0.728 and Compression-Aware Information Accuracy from 1.327 to 1.511. On non-ML scientific papers, the gains are similarly consistent, with Compression-Aware Information Accuracy increasing from 1.339 to 1.611 and Aesthetic Score from 2.53 to 2.66. These results suggest that AutoPage is not tied to ML-specific writing conventions or figure styles, but generalizes well across broader technical and scientific documents through structure-aware parsing and hierarchical planning.

\section{Details of PageBench}
\label{sec:appendix_pagebench}

\subsection{Readability}
\label{sec:appendix_pagebench_read}

Perplexity (PPL) has traditionally been regarded as a classic indicator of linguistic fluency, and many early works on automatic text adopt it as a proxy for readability~\cite{pang2025paper2poster}. However, PPL fundamentally measures the predictability of a token sequence under a language model, making it sensitive to domain-specific terminology, mathematical expressions, and other non-colloquial constructs. As a result, PPL often diverges from human judgments of readability on academic or technical webpages.

Recent systematic analysis~\cite{cachola2025evaluatingevaluatorsreadabilitymetrics} further shows that traditional surface-level readability metrics correlate poorly with human assessments, especially on technical texts. In contrast, LLM-as-a-judge approaches are shown to better capture discourse-level clarity, coherence, and organizational quality. 
Motivated by these findings and by the limitations of PPL in our evaluation domain, we adopt an LLM-based readability assessment framework to obtain more stable and human-aligned readability judgments.

\subsection{Semantic Fidelity}

\label{sec:appendix_pagebench_semantic}

The Semantic Fidelity score quantifies the preservation of meaning between the generated webpage content and the source document.
Although BERTScore~\cite{zhang2020bertscoreevaluatingtextgeneration} is effective for token-level tasks such as summarization, it is often overly sensitive to structural rephrasing (\textit{e.g.}~merging, splitting) common in complex webpage generation. For this reason, we adopt an approach based on paragraph-level semantic alignment, which is more robust to high-level content restructuring.

Inspired by similar recent work~\cite{guo2025paper2sysarchstructureconstrainedarchitecturegeneration}, we use Sentence-BERT~\cite{reimers2019sentencebertsentenceembeddingsusing} to measure Semantic Fidelity. Our calculation is a multi-stage process designed to be robust and accurate. 
Given that generated webpages may involve content merging, splitting, or rephrasing, we first align each generated section to its most semantically similar source paragraph using sentence embeddings, yielding a set of aligned (generated section, source paragraph) pairs.
For each aligned pair, we use a pre-trained sentence-transformer model from huggingface\footnote{\href{https://huggingface.co/sentence-transformers/all-roberta-large-v1}{https://huggingface.co/sentence-transformers/all-roberta-large-v1}} to encode both the generated text and the source text into high-dimensional vectors, denoted as $V_g$ and $V_s$ respectively. These dense vectors capture the semantic meaning of the text.
We then compute the cosine similarity between the two vectors $V_g$ and $V_s$. This value, ranging from -1 to 1, measures the cosine of the angle between them. A value close to 1 indicates that the vectors point in almost the same direction, signifying a high degree of semantic similarity. The formula is:
\begin{equation}
\text{Semantic Similarity}(V_g, V_s) = \frac{V_g \cdot V_s}{\|V_g\| \|V_s\|}
\end{equation}
The final Semantic Fidelity score for the entire webpage is the average of the cosine similarity scores across all aligned section pairs. This provides a single, comprehensive measure of how well the webpage preserves the semantics of the source document as a whole.

\begin{figure}[!t]
            \centering
            \includegraphics[width=\linewidth]{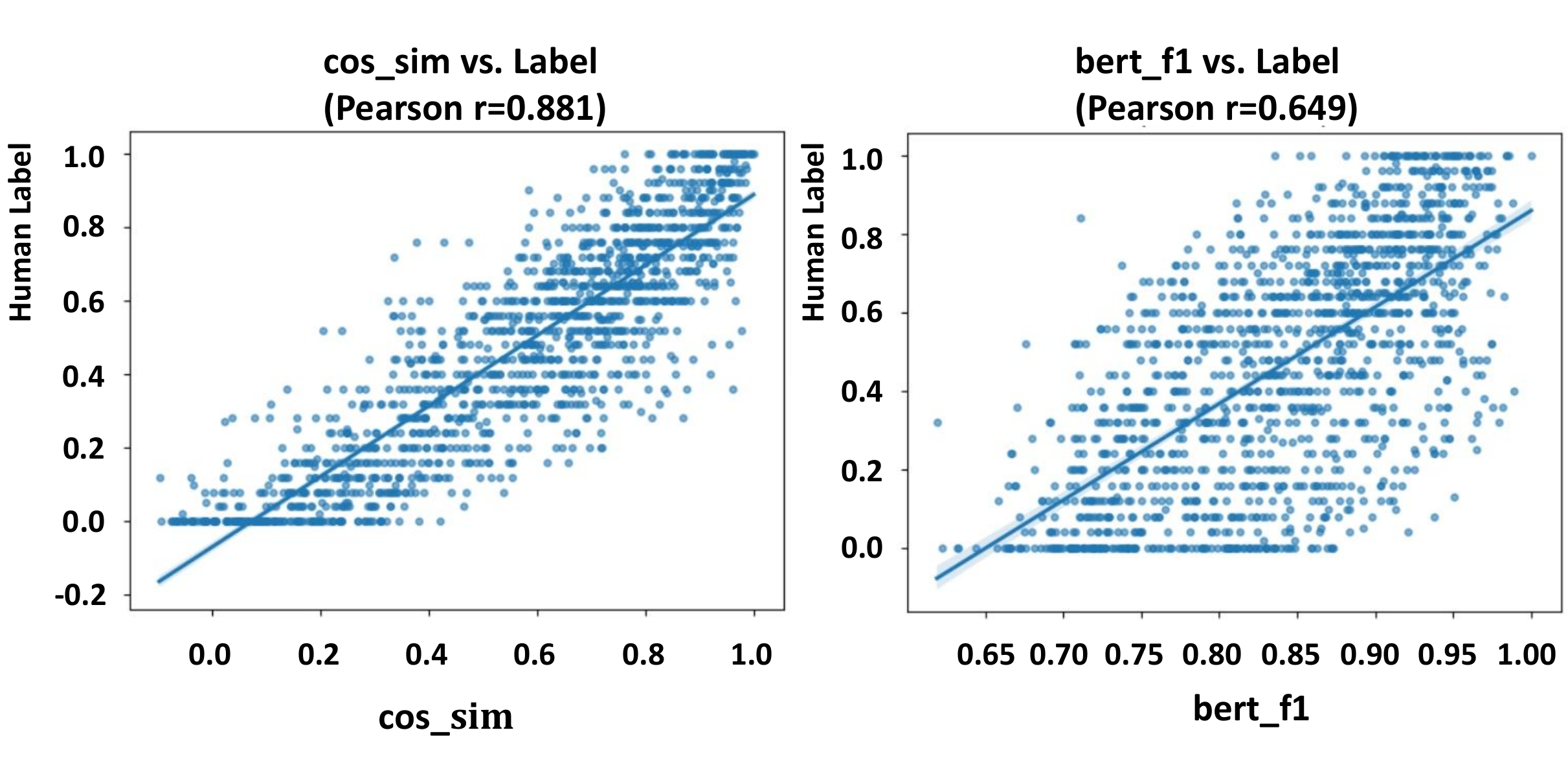}
            \vspace{-15pt}
            \caption{\textbf{STS-B Validation.} Pearson correlation with human semantic similarity scores on STS-B~\cite{Cer_2017}. The sentence-level cosine similarity metric based on Sentence-BERT (left) achieves a strong correlation~(\textbf{r=0.881}), substantially outperforming the token-level BERTScore F1 baseline~(right, \textbf{r = 0.649}).
            This indicates that sentence embedding–based similarity is more robust to paraphrasing and structural transformations, making it better suited for paragraph-level semantic fidelity evaluation.}
     \label{fig:cos_bert}
     \vspace{-10pt}
\end{figure}

To confirm the discriminative ability of our metric, we validated its correlation with human judgment using the Semantic Textual Similarity Benchmark (STS-B) dataset~\cite{Cer_2017}. 
As shown in Fig.~\ref{fig:cos_bert}, the sentence-level cosine similarity metric based on Sentence-BERT exhibits a strong Pearson correlation with human semantic similarity judgments on the STS-B benchmark. This result supports its use as a robust measure of semantic fidelity, particularly for evaluation settings involving paraphrasing and structural reorganization.

\subsection{Compression-Aware Information Accuracy}
\label{sec:appendix_pagebench_qa}
\begin{table*}[hbtp]
\centering
\caption{\textbf{Performance Comparison across Different Methods.} The best performance among all methods for each metric is in \textbf{bold}, and the second best is \underline{underlined}. For ease of comparison, AutoPage and its corresponding proprietary base models are highlighted in matching colors. The compression rate is also listed in the table. }
\vspace{-10pt}
\label{tab:qa_performance}
\resizebox{\textwidth}{!}{
\begin{tabular}{lcc|ccc|ccc|c|c|ccc}
\toprule
\multirow{3}{*}{\textbf{Method}} & \multirow{3}{*}{\textbf{System Type}} & \multirow{3}{*}{\textbf{Open-Source}} & \multicolumn{7}{c|}{\textbf{Raw-ACC}} & \multirow{3}{*}{\textbf{Compression}} & \multicolumn{3}{c}{\textbf{Compression-Aware ACC}} \\
\cmidrule(lr){4-10} \cmidrule(lr){12-14}
 &  &  & \multicolumn{3}{c|}{\textbf{Detail question}} & \multicolumn{3}{c|}{\textbf{Understanding question}} & \multirow{2}{*}{\textbf{Overall}$\uparrow$} &  & \multirow{2}{*}{\textbf{D-Avg}$\uparrow$} & \multirow{2}{*}{\textbf{U-Avg}$\uparrow$} & \multirow{2}{*}{\textbf{Overall$\uparrow$}} \\
\cmidrule(lr){4-6} \cmidrule(lr){7-9}
 &  &  & \begin{tabular}[c]{@{}c@{}}Open-\\Source\end{tabular} & \begin{tabular}[c]{@{}c@{}}Close-\\Source\end{tabular} & \begin{tabular}[c]{@{}c@{}}\textbf{D-Avg}$\uparrow$\end{tabular} & \begin{tabular}[c]{@{}c@{}}Open-\\Source\end{tabular} & \begin{tabular}[c]{@{}c@{}}Close-\\Source\end{tabular} & \begin{tabular}[c]{@{}c@{}}\textbf{U-Avg}$\uparrow$\end{tabular} &  &  &  &  \\
\midrule
\quad GPT-OSS-120B & E2E & \textcolor{gray}{\usym{2714}} & 0.662 & 0.617 & 0.640 & 0.869 & 0.844 & 0.856 & 0.748 & 10.833 & 1.469 & 1.970 & 1.719 \\
\quad llama-3.1-70B & E2E & \textcolor{gray}{\usym{2714}} & 0.422 & 0.218 & 0.320 & 0.617 & 0.380 & 0.499 & 0.409 & 29.594 & 0.992 & 1.548 & 1.270 \\
\quad Grok-4-fast & E2E &  & \underline{0.725} & 0.711 & \underline{0.718} & 0.880 & 0.890 & \textbf{0.885} & \underline{0.801} & 10.347 & 1.617 & 1.999 & 1.808 \\
\quad GLM-4.5-Air & E2E &  & 0.688 & 0.619 & 0.653 & \underline{0.905} & 0.850 & \underline{0.877} & 0.765 & 11.213 & 1.526 & 2.049 & 1.788 \\
\midrule
\quad Qwen3-235B-A22B & E2E & \textcolor{gray}{\usym{2714}} & 0.692 & 0.547 & 0.620 & 0.859 & 0.794 & 0.826 & 0.723 & 15.931 & \textbf{1.659} & \underline{2.121} & \underline{1.890} \\
\rowcolor{lemonchiffon}\quad AutoPage-Qwen & Multi-Agent &  & 0.700 & 0.635 & 0.668 & \textbf{0.914} & 0.833 & 0.874 & 0.771 & 11.592 & 1.593 & 2.081 & 1.837 \\
\midrule
\quad GPT-4o-mini & E2E &  & 0.635 & 0.374 & 0.505 & 0.838 & 0.606 & 0.722 & 0.613 & 20.045 & 1.472 & 2.099 & 1.786 \\
\rowcolor{lightcoral}\quad AutoPage-GPT-4o-mini & Multi-Agent &  & 0.635 & 0.398 & 0.517 & 0.899 & 0.662 & 0.781 & 0.649 & 23.528 & 1.581 & \textbf{2.389} & \textbf{1.941} \\
\midrule
\quad Gemini-2.5-flash & E2E &  & \textbf{0.735} & \underline{0.723} & \textbf{0.729} & 0.869 & \textbf{0.901} & \textbf{0.885} & \textbf{0.807} & 5.302 & 1.150 & 1.402 & 1.276 \\
\rowcolor{skyblue}\quad AutoPage-Gemini-2.5-flash & Multi-Agent &  & 0.715 & 0.687 & 0.701 & 0.882 & 0.870 & 0.876 & 0.788 & 8.419 & 1.411 & 1.770 & 1.591 \\
\bottomrule

\quad GPT-5-mini & E2E &  & 0.637 & 0.662 & 0.649 & 0.801 & 0.849 & 0.825 & 0.737 & 10.043 & 1.4 & 1.778 & 1.589 \\
\rowcolor{SeaGreen}\quad AutoPage-GPT-5-mini & Multi-Agent &  & 0.701 & \textbf{0.728} & 0.715 & 0.816 & \underline{0.891} & 0.853 & 0.784 & 11.962 & \underline{1.658} & 1.978 & 1.818 \\
\bottomrule
\end{tabular}
}
\vspace{-15pt}
\end{table*}

The Compression-Aware Information Accuracy metric is designed to jointly evaluate the factual accuracy and conciseness of the generated content. The calculation involves the following four steps:

\paragraph{QA-Based Accuracy Measurement.}
We first automatically generate a set of 100 question-answer pairs from the source document using a powerful large language model \texttt{GPT-o3}. Similar to Paper2Poster~\cite{pang2025paper2poster}, we select six powerful large language models to answer the questions above, including GPT-4o-mini~\cite{achiam2023gpt}, gemini-2.5-flash~\cite{team2023gemini}, grok-4-fast~\cite{xGrok}, Phi4-14B~\cite{abdin2024phi}, Mistral-small-3.1-24B~\cite{mistralMistralSmall} and Qwen3-14B~\cite{qwen3}. Then, an answering model is tasked to answer these questions based solely on the textual content extracted from the generated webpage.
The QA Accuracy, denoted as $A$, is the fraction of correctly answered questions.

\paragraph{Text Compression Ratio.}
The Text Compression Ratio, $C$, measures how much shorter the generated text is compared to the original source text. It is defined as the ratio of the token counts:
\begin{equation}
C = \frac{\text{Tokens}_{\text{ori}}}{\text{Tokens}_{\text{gen}}}
\end{equation}
A value of $C > 1$ indicates compression.

\paragraph{Final Score Calculation.}
To combine accuracy and compression into a single score, $S_{\text{final}}$, we multiply the accuracy by the natural logarithm of the compression ratio. Using the logarithm rewards conciseness while dampening the effect of extreme compression. The formula is:
\begin{equation}
S_{\text{final}} = A \times \ln(C)
\end{equation}
This final, normalized score effectively and comparably rewards models that produce concise yet factually accurate content.


\subsection{VLM-as-Judge for Visual Quality Evaluation}
\label{sec:appendix-vlm-as-judge}
To quantitatively assess the visual quality of the generated pages, we employ a Vision Language Model (VLM) as an automated judger.
This approach, termed \textit{``VLM-as-Judge''}~\cite{liu2023mitigating,liu2023visual,zhu2023judgelm,lee2024prometheusvision,gao2025rapo++,ma2025led,Ma_2025_CVPR,Lan_2025_CVPR,huang2025vbench++}, allows for a consistent and scalable evaluation of the key visual elements presented in the documents, focusing specifically on their correctness and relevance rather than on subjective aesthetic appeal.
The VLM is guided by a carefully designed system prompt, instructing it to act as an extremely strict visual elements reviewer. Detailed prompt templates for visual quality evaluation as listed in Appendix~\ref{sec:appendix-prompt-templates}.

\section{Detailed Analysis of Compression-Aware QA Performance}
\label{sec:appendix-comp-acc}
A comprehensive breakdown of the Question Answering (QA) results is provided in Tab.~\ref{tab:qa_performance}. We report performance using two primary metrics: Raw Accuracy and Compression-Aware Accuracy, the latter of which is modulated by the compression ratio, as described in Appendix~\ref{sec:appendix_pagebench_qa}.
The results are further disaggregated by question type. Our analysis yields several key observations:
(i)~AutoPage variants significantly enhance their base models' performance. For instance, AutoPage-GPT-4o-mini elevates its base model's score from 1.786 to 1.941, while AutoPage-Gemini-2.5-flash improves its score from 1.276 to 1.591, validating our structured compression method.
(ii)~While leading end-to-end models excel in raw accuracy, AutoPage demonstrates a distinct advantage by achieving much higher compression rates while maintaining comparable accuracy. This superiority is reflected in the Compression-Aware ACC metric, where AutoPage-GPT-4o-mini attains the highest overall score of 1.941, underscoring the efficacy of our multi-agent approach in efficient information distillation.
In summary, AutoPage proves its value by achieving significantly higher compression rates than E2E models while delivering comparable raw accuracy. This effective balance sets a new benchmark in compression-aware evaluations.

\section{Details for the Test Set and Template Library Construction}
\label{sec:sampling_test_template}
To construct a diverse and representative test set, we moved beyond random sampling. We implemented a two-stage diversity sampling strategy, designed to produce two distinct assets: a diverse Test Set and a stylistically unique Template Library.

\noindent First, to create the test set, we performed feature extraction on the entire corpus to capture its structural and stylistic properties. We then applied dimensionality reduction and clustering techniques to group pages with similar layouts. By sampling from the resulting clusters, we selected approximately 100 pages that represent a broad range of the page archetypes found in the wild. This collection serves as our primary test set for evaluation.
Second, to build a library of unique design patterns for template matching tasks, we further refined this test set. We applied a multi-stage deduplication algorithm to the 100 candidate pages. This algorithm combines rapid SimHash-based filtering with a precise Zhang-Shasha~\cite{Zhang1989SimpleFA} tree edit distance computation on standardized DOM structures to identify and remove pages originating from the same template. The most structurally complex page from each identified group was chosen as the representative. This filtering step yielded a final, curated collection of 87 stylistically distinct pages, which constitutes our Template Library. 

\begin{figure}[!t]
            \centering
            \includegraphics[width=\linewidth]{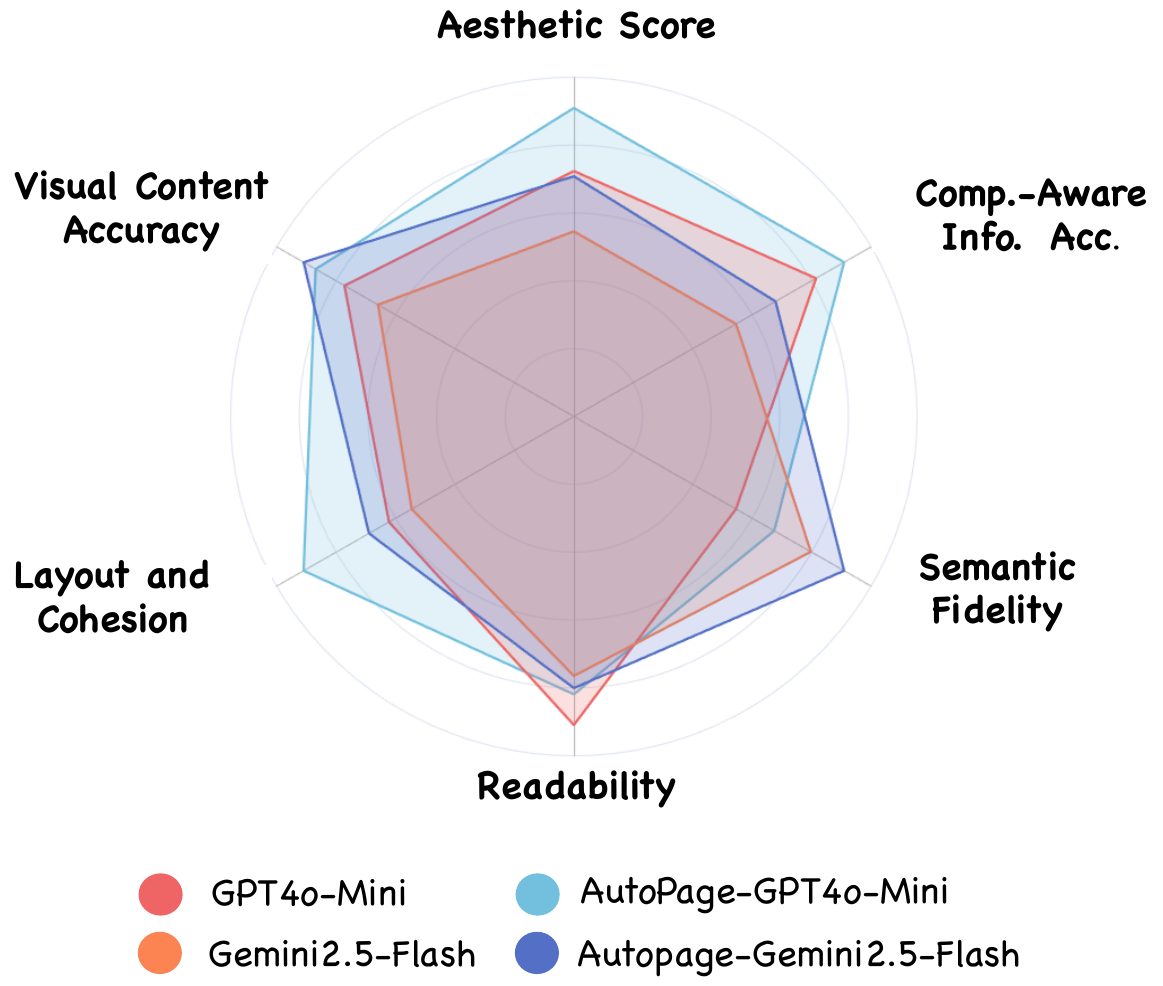}

            \caption{\textbf{Comprehensive Evaluation of Model Performance with and without AutoPage.} 
            The radar plot shows that integrating AutoPage consistently boosts both content and visual quality, thereby demonstrating its strong advantage in generating more accurate, coherent, and visually appealing webpages.}
            
            
     \label{fig:radar-chart}

\end{figure}

\section{User Study Protocol and Materials}
\label{sec:appendix-user-study}
This section provides the detailed protocol, recruitment materials, and scoring rubric used in our user study, as referenced in the main paper.

\paragraph{Participants and Recruitment.}
We recruited 20 undergraduate and graduate students to act as expert evaluators. Participants were informed that the task would take approximately 2 hours and would be compensated upon successful completion. Each participant was compensated accordingly for their contribution.

\paragraph{Task Procedure and Instructions.}
Participants were given a detailed guide explaining their task. For each set, they were instructed to evaluate 8 webpages generated by different AI models and our system from the same source paper. The instructions guided them to first quickly browse all 8 pages to form a general impression and a preliminary mental ranking. Following this, they were to begin the scoring process by assigning a high score (e.g., 10 or 9) to the best page and a low score (e.g., 1 or 2) to the worst, establishing anchors for their judgment. Finally, they would assign the remaining pages unique, unused scores based on their quality relative to these anchors and to each other. Once all 8 pages in a group were scored, the results were submitted, and the system would present the next group.

\paragraph{Forced-Choice Scoring Rubric.}
The core of our study was the forced-choice rating system. Participants were required to use a unique integer score from 1 to 10 for each page within a single group of 8. The detailed rubric provided to them was as follows:
\begin{table*}[t]
\centering
\caption{\textbf{Absolute Quality Ratings (5-point Likert Scale).} 
Participants rated each generated webpage along five dimensions: readability, content quality, multimodal rendering correctness, layout coherence, and visual aesthetics. AutoPage achieves the highest score across all dimensions.}
\vspace{-6pt}
\label{tab:likert_results}
\resizebox{\textwidth}{!}{
\begin{tabular}{lcccccccc}
\toprule
\textbf{Dimension / Model} 
& \textbf{GPT4o-mini}
& \textbf{Gemini2.5-Flash}
& \textbf{Grok4-fast}
& \textbf{GLM-4.5-Air}
& \textbf{gpt-oss-120b}
& \textbf{llama-3.1-70b}
& \textbf{qwen3-235b-a22b}
& \textbf{AutoPage} \\
\midrule
Readability
& 3.18$\pm$0.25 
& 3.63$\pm$0.34
& 3.59$\pm$0.40
& 3.89$\pm$0.37
& 3.49$\pm$0.46
& 2.95$\pm$0.47
& 3.17$\pm$0.47
& \textbf{3.90$\pm$0.25} \\
Content Quality
& 3.09$\pm$0.25 
& 3.54$\pm$0.46
& 3.68$\pm$0.44
& 3.68$\pm$0.25
& 3.45$\pm$0.26
& 2.76$\pm$0.46
& 2.97$\pm$0.48
& \textbf{3.89$\pm$0.23} \\
Multimodal Render
& 2.41$\pm$0.51 
& 3.45$\pm$0.38
& 3.49$\pm$0.47
& 3.34$\pm$0.38
& 3.26$\pm$0.38
& 2.34$\pm$0.31
& 2.73$\pm$0.49
& \textbf{3.92$\pm$0.22} \\
Layout \& Cohesion
& 2.52$\pm$0.50 
& 3.51$\pm$0.35
& 3.54$\pm$0.35
& 3.14$\pm$0.38
& 3.31$\pm$0.38
& 2.15$\pm$0.29
& 2.67$\pm$0.49
& \textbf{3.93$\pm$0.26} \\
Visual Aesthetics
& 2.50$\pm$0.47 
& 3.60$\pm$0.37
& 3.60$\pm$0.45
& 3.10$\pm$0.37
& 3.26$\pm$0.37
& 2.55$\pm$0.29
& 2.75$\pm$0.49
& \textbf{3.91$\pm$0.24} \\
\midrule
\textbf{Average}
& 2.73$\pm$0.19 
& 3.54$\pm$0.22
& 3.58$\pm$0.24
& 3.43$\pm$0.15
& 3.31$\pm$0.19
& 2.55$\pm$0.17
& 2.87$\pm$0.22
& \textbf{3.91$\pm$0.10} \\
\bottomrule
\end{tabular}
}
\vspace{-10pt}
\end{table*}

\begin{itemize}
    \item \textbf{10 (Perfect):} Professional, flawless design and content presentation. A "gold standard" page.
    \item \textbf{9 (Excellent):} Near-perfect, with only minuscule flaws discoverable upon close inspection.
    \item \textbf{8 (Good):} High-quality page with complete functionality, but perhaps lacking in design flair.
    \item \textbf{7 (Decent):} Generally usable but with minor, noticeable issues like misalignment or incorrect rendering of non-critical content.
    \item \textbf{6 (Fair/Average):} Inconsistencies or minor clutter in layout and content, but does not impede usability.
    \item \textbf{5 (Marginally Usable):} Noticeable design issues (e.g., slight element overlap) that begin to affect the reading experience.
    \item \textbf{4 (Poor):} Chaotic layout with significant content rendering failures, severely hindering comprehension of core content.
    \item \textbf{3 (Very Poor):} The layout is mostly broken (akin to failed CSS), making most content unreadable.
    \item \textbf{2 (Broken):} The page is fundamentally broken, with only scattered, isolated pieces of content being recognizable.
    \item \textbf{1 (Completely Unusable):} A blank page, a server/browser error (e.g., 404), or complete gibberish. The page has zero value.
\end{itemize}

\paragraph{Why We Adopt Forced-Choice?}

The forced-choice design was adopted to obtain stable relative preferences across webpage variants. In preliminary pilot studies, we observed substantial variability in how participants used absolute rating scales, which made direct comparison of raw numeric scores across annotators unreliable. Forced-choice judgments, in contrast, reduce scale-use bias and directly capture the relative quality differences that are central when comparing multiple generation systems. This protocol also helps mitigate order effects: participants must view all webpages before assigning scores, reducing the tendency to anchor early webpages with disproportionately high or low ratings and encouraging more consistent, globally informed judgments. Because forced-choice scoring produces a ranking rather than an absolute quality measure, we additionally conducted a supplementary rating-based study using a 5-point Likert scale. Participants independently rated each webpage along five quality dimensions: readability, content quality, multi-modal rendering correctness, layout coherence, and visual aesthetics. This evaluation provides absolute, dimension-level judgments that complement the relative comparisons of the forced-choice protocol. 

\noindent As shown in Tab.~\ref{tab:likert_results} the Likert results strongly align with the forced-choice outcomes and consistently show that our system achieves the highest scores across all dimensions.  The strong agreement between the two evaluation paradigms demonstrates that forced-choice scoring faithfully captures human preferences and that the relative advantages observed for our system are robust across both ranking-based and absolute rating-based assessments.

\section{Correlation with human judgments.}

Beyond the human preference study in Figure~\ref{fig:user-study}, we further examine whether our automatic metrics are aligned with human judgments. Using the rating samples collected in the user study, we compute Kendall's $\tau$ rank correlation\cite{10.1093/biomet/30.1-2.81} between automatic scores and human ratings for each evaluated dimension. As shown in Table~\ref{tab:human_corr}, all dimensions exhibit positive correlations. In particular, \textit{Readability} shows a relatively strong correlation of 0.69, suggesting that our automatic evaluation captures human perception of discourse clarity and structural quality reasonably well. The correlations for \textit{Visual Content Accuracy} (0.40), \textit{Layout and Cohesion} (0.36), and \textit{Aesthetic Score} (0.22) are more moderate, but still indicate consistent alignment trends between automatic metrics and human preferences.
\begin{table}[t]
\centering
\caption{Kendall's $\tau$ rank correlation between automatic metrics and human judgments.}
\label{tab:human_corr}
\scriptsize
\setlength{\tabcolsep}{3.5pt}
\renewcommand{\arraystretch}{1.08}
\begin{tabular}{lcccc}
\toprule
& \textbf{Readability}
& \makecell[c]{\textbf{Visual Content}\\\textbf{Acc.}}
& \makecell[c]{\textbf{Layout \&}\\\textbf{Cohesion}}
& \makecell[c]{\textbf{Aesthetic}\\\textbf{Score}} \\
\midrule
Kendall's $\tau$ & 0.69 & 0.40 & 0.36 & 0.22 \\
\bottomrule
\end{tabular}
\end{table}
\section{Discussion}
\label{sec:discussion}

\begin{table}[h!]
\centering
\caption{\textbf{Generation time and monetary cost of AutoPage compared to manual creation.}}
\vspace{-6pt}
\label{tab:efficiency_cost}

\renewcommand{\arraystretch}{1.15}

\resizebox{\linewidth}{!}{
{\large
\begin{tabular}{lcc}
\toprule
\textbf{Method} & \textbf{Cost (USD)} & \textbf{Time (min)} \\
\midrule
Human (Manual Creation) & --   & \textbf{120} \\
\midrule
AutoPage--Qwen              & 0.14 & 15 \\
AutoPage--GPT-4o-mini       & 0.07 & 12 \\
AutoPage--Gemini-2.5-Flash  & 0.20 & 6.25 \\
\bottomrule
\end{tabular}
}}
\vspace{-10pt}
\end{table}

\noindent\textbf{Efficiency and cost-effectiveness analysis.}
A brief survey of 10 PhD researchers indicates that manually creating a project webpage typically requires about \textbf{2 hours}. In contrast, AutoPage generates a full page in \textbf{6--15 minutes} at a cost of only \textbf{\$0.06--\$0.20}, depending on the chosen model.

\noindent This shows that AutoPage delivers high-quality webpages while reducing human effort by an order of magnitude with minimal monetary overhead.

This remarkable cost-effectiveness confirms that AutoPage is not merely a proof-of-concept, but a scalable and economically viable solution ready for real-world deployment.

\section{Additional Comparision Resuls}
\begin{figure*}[!t]
            \centering
            \includegraphics[width=0.85\textwidth]{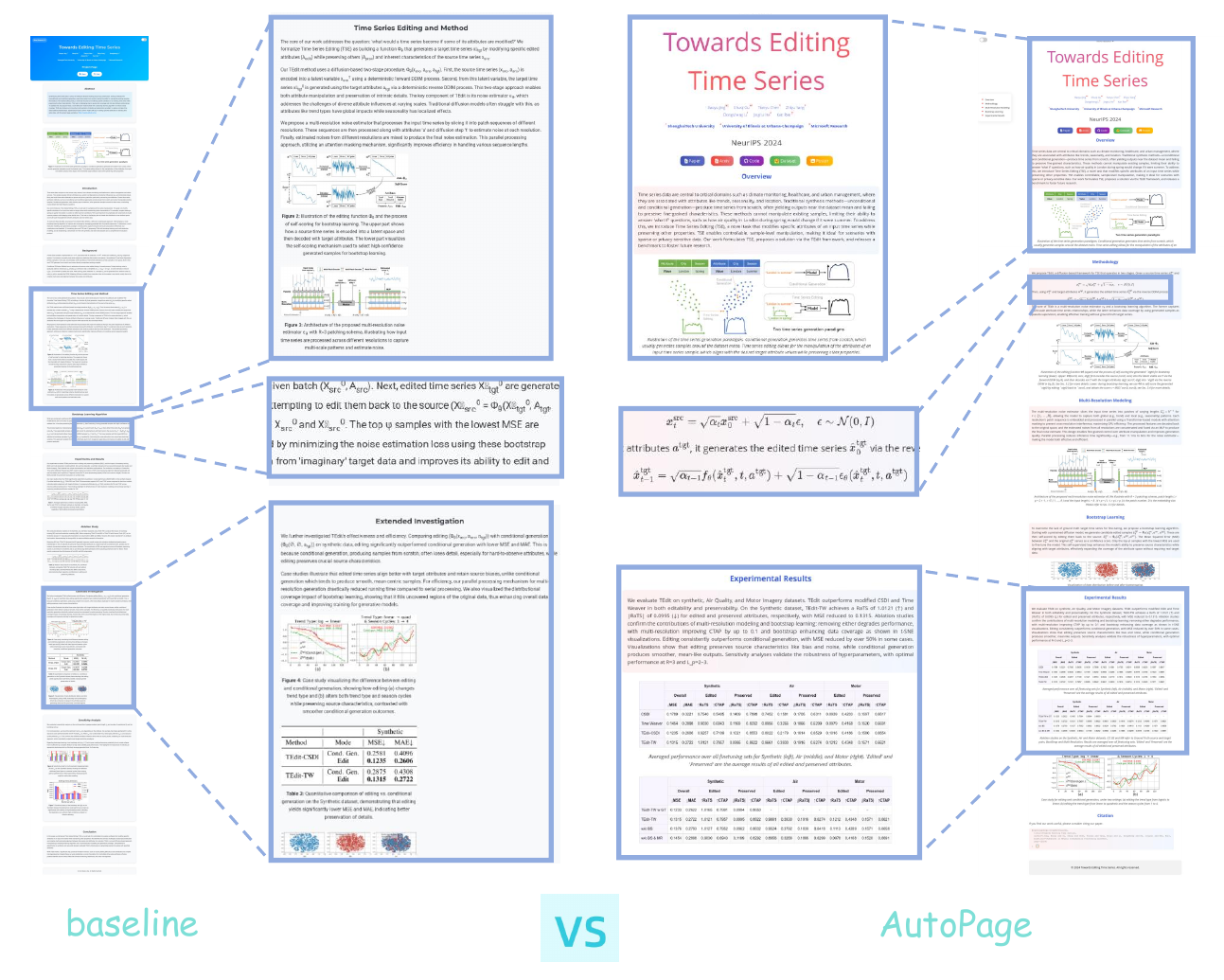}
            \caption{\textbf{Visual comparison of baseline and AutoPage}. The webpage generated by the baseline (left) exhibits rendering failures, including an inability to display the formula and a distorted layout for images and tables. Conversely, the page generated by AutoPage (right) renders the formula correctly and preserves the intended layout of all visual elements.}
     \label{fig:app-visual_results1}
    \vspace{-10pt}
\end{figure*}

\begin{figure*}[!t]

            \centering
            \includegraphics[width=0.82\textwidth]{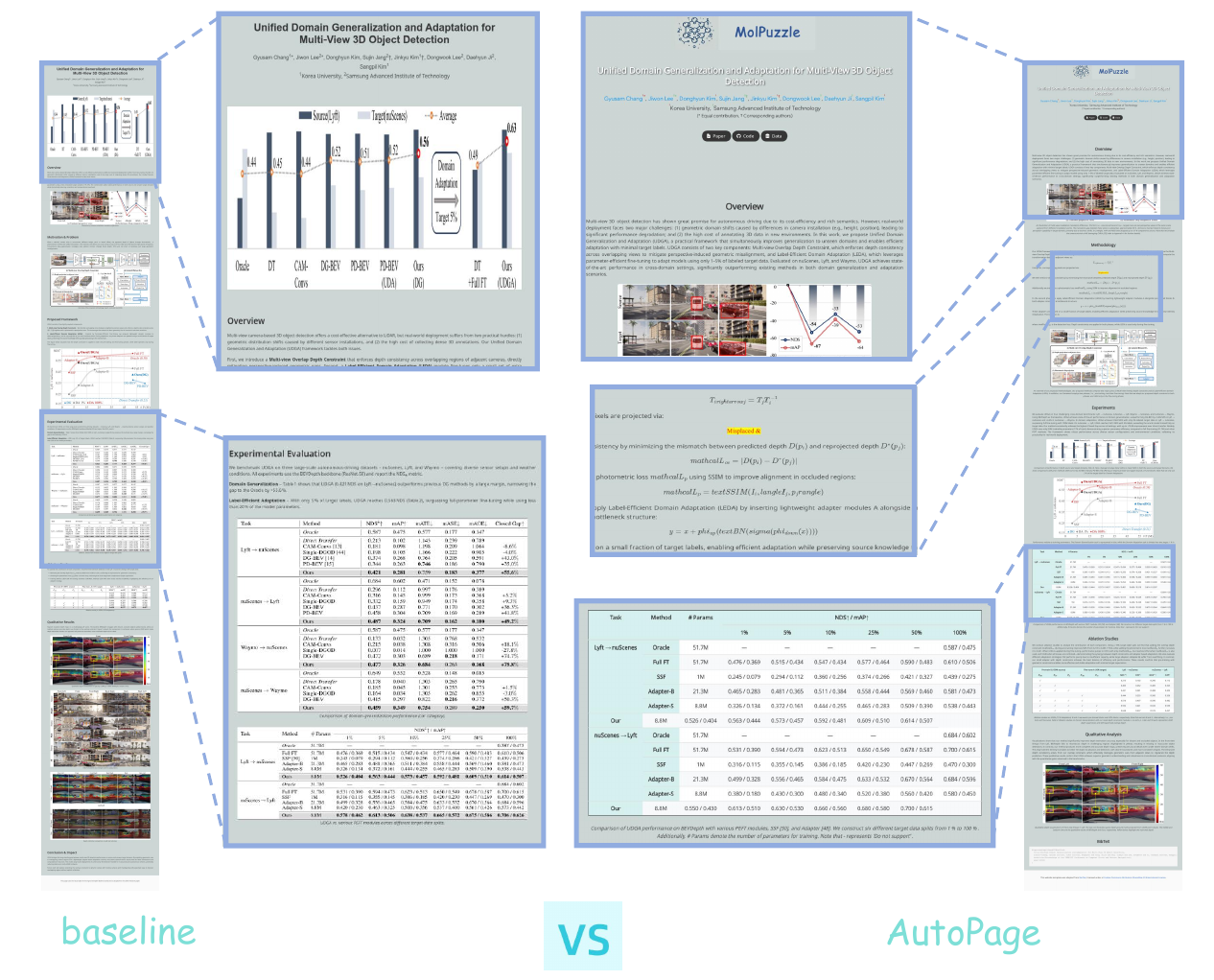}
            \caption{\textbf{Additional Visual comparison of baseline and AutoPage}. The baseline-generated page (left) suffers from improper image scaling, leading to a catastrophic visual presentation. In contrast, AutoPage (right) styles the tables to match the page's theme and displays the image at an optimal size.}
     \label{fig:app-visual_results2}
    \vspace{-10pt}
\end{figure*}

\begin{figure*}[!t]

            \centering
            \includegraphics[width=0.82\textwidth]{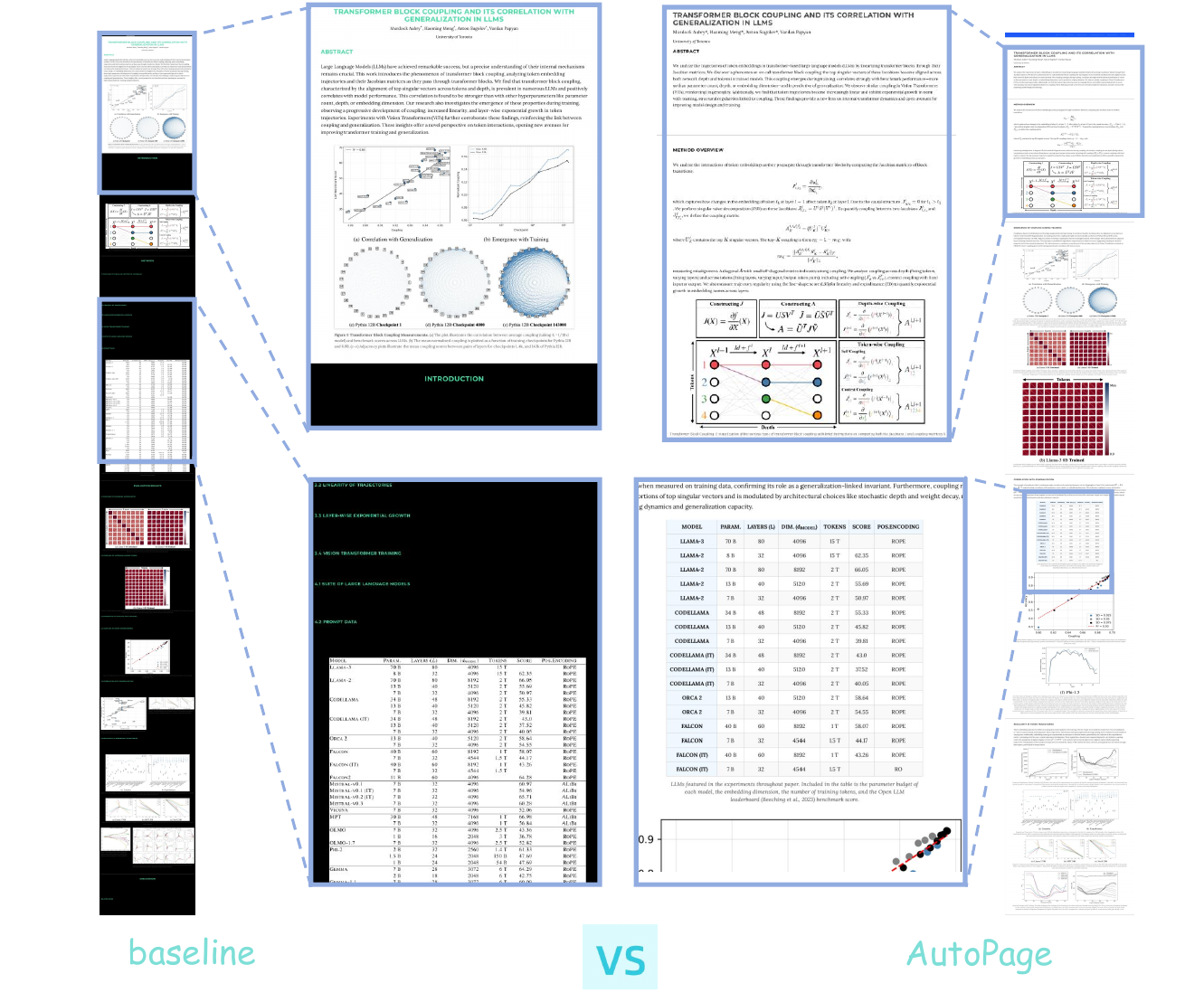}
            \caption{\textbf{Additional Visual comparison of baseline and AutoPage}. The webpage generated by the baseline method (left) fails to render the contents within their respective sections. Conversely, the page generated by AutoPage (right) successfully displays both the textual and visual elements in their correct layout.}
     \label{fig:app-visual_results3}
    \vspace{-10pt}
\end{figure*}

\begin{figure*}[!t]
            \centering
            \includegraphics[width=1.0\textwidth]{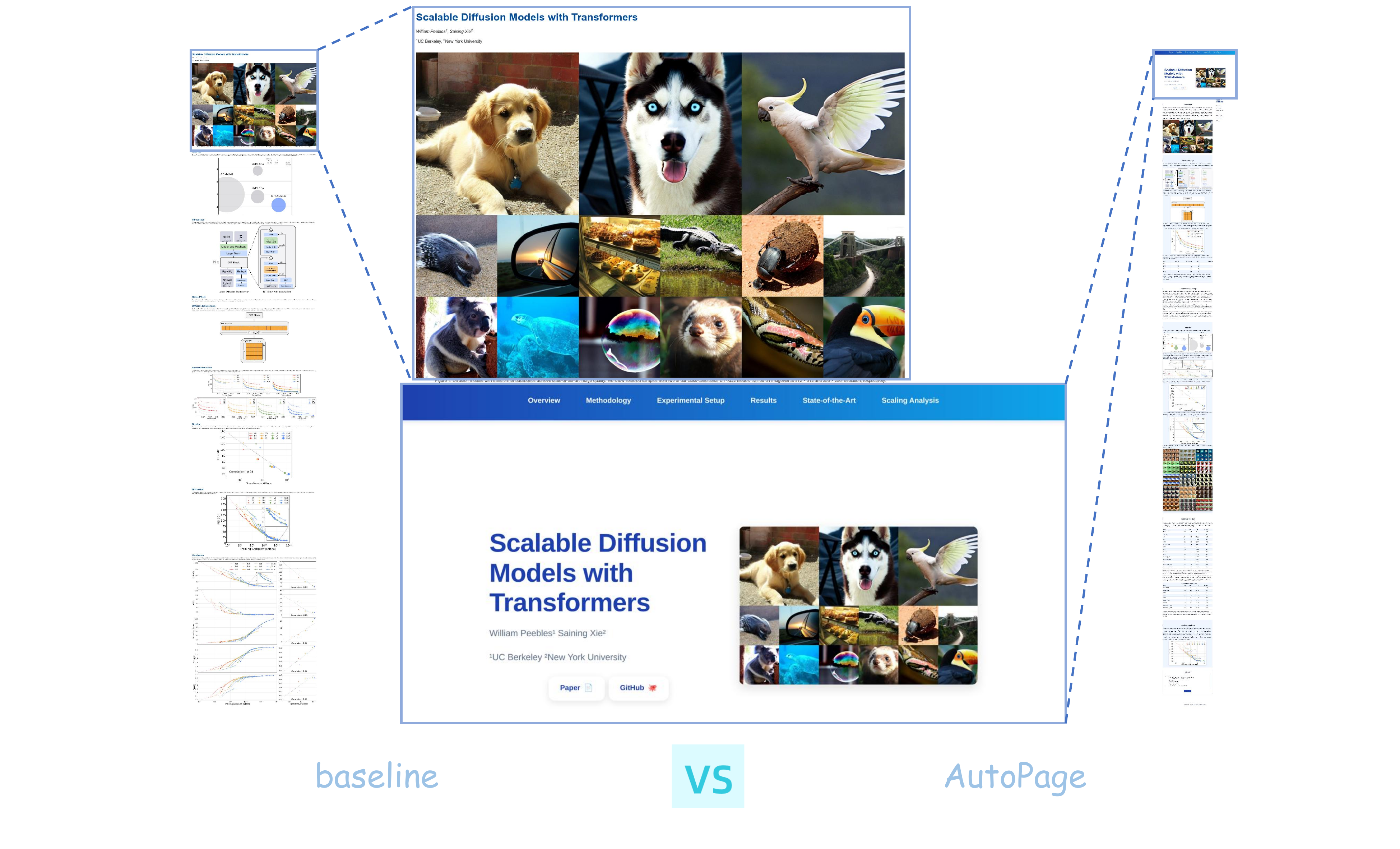}
            \caption{\textbf{Template-free visual comparison of baseline and AutoPage}.The webpage generated by the baseline (left) exhibits a pronounced aspect-ratio issue in the hero section, where the cover image is improperly stretched and/or cropped, leading to a distorted visual appearance and misaligned composition. In contrast, the page generated by AutoPage (right) preserves the correct image aspect ratio with appropriate scaling/cropping in the hero section, yielding a more natural and visually consistent presentation.}
     \label{fig:Template-free-visual_results1}
    \vspace{-10pt}
\end{figure*}

\begin{figure*}[!t]
            \centering
            \includegraphics[width=1.0\textwidth]{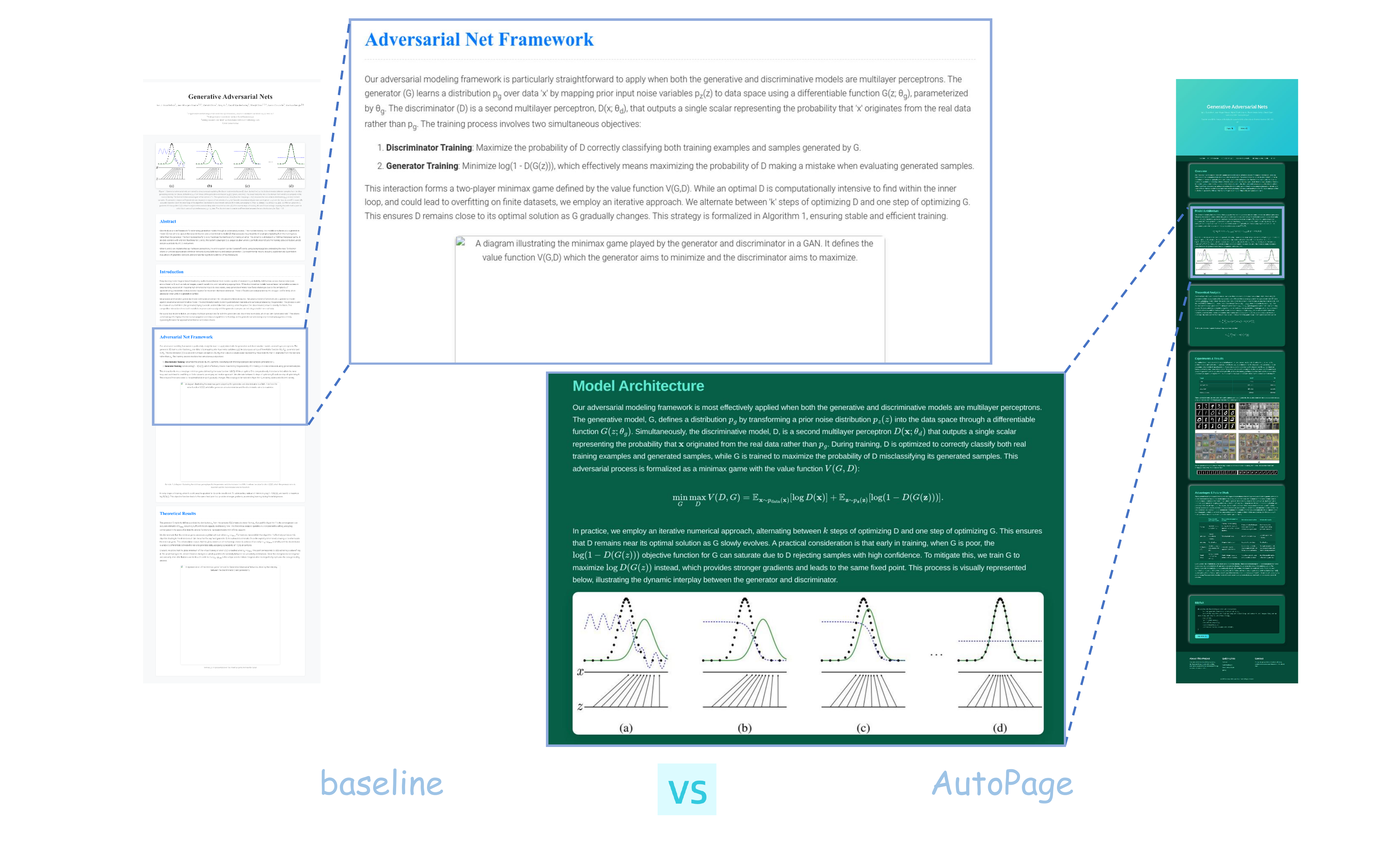}
            \caption{\textbf{Template-free visual comparison of baseline and AutoPage}. The webpage generated by the baseline (left) exhibits a content-planning failure where key formulas are not properly incorporated into the page structure, causing crucial derivations to be omitted; it also suffers from missing images due to incorrect image paths/URLs. In contrast, the page generated by AutoPage (right) successfully integrates key formulas into the appropriate sections during content planning and correctly resolves image resources, resulting in a more complete and faithful webpage rendering of the paper.}
     \label{fig:Template-free-visual_results2}
    \vspace{-10pt}
\end{figure*}

In this section, we present additional visual results in Fig.~\ref{fig:app-visual_results1}, Fig.~\ref{fig:app-visual_results2}, Fig.~\ref{fig:app-visual_results3}, Fig.~\ref{fig:Template-free-visual_results1} and Fig.~\ref{fig:Template-free-visual_results2} to further demonstrate the superiority of AutoPage over the baseline method, including comparisons under both template-free and template-based settings for the baseline and AutoPage.

\noindent As shown in Fig.~\ref{fig:app-visual_results1}, the baseline method struggles with complex page elements, failing to render mathematical formulas and disrupting the layout of images and tables. In contrast, AutoPage accurately renders all components, preserving the page's intended structure.

\noindent Fig.~\ref{fig:app-visual_results2} highlights the difference in visual fidelity. The baseline's improper image scaling leads to a catastrophic visual presentation. AutoPage, however, not only resizes images appropriately but also adapts table styles to the page's theme, creating a cohesive and aesthetically pleasing result.

\noindent Fig.~\ref{fig:app-visual_results3} reveals a more fundamental failure of the baseline: content loss. The baseline-generated page fails to display content within entire sections, rendering it incomplete. AutoPage, conversely, ensures content integrity by correctly displaying all textual and visual elements.

\noindent Fig.~\ref{fig:Template-free-visual_results1} emphasizes the robustness of AutoPage in handling the hero section. The baseline-generated webpage exhibits severe aspect-ratio issues, where the hero image is improperly stretched or cropped, resulting in an unnatural and visually unbalanced presentation. AutoPage, however, preserves the correct image aspect ratio and applies appropriate scaling/cropping, producing a visually faithful hero layout that remains consistent with the overall page design.

\noindent Finally, Fig.~\ref{fig:Template-free-visual_results2} reveals a more fundamental limitation of the baseline method: content planning and resource resolution failures. Key formulas are not properly incorporated into the generated page structure, causing crucial derivations to be omitted, and incorrect image paths further lead to missing visual assets. In contrast, AutoPage integrates key formulas into their appropriate sections during content planning and correctly resolves image resources, ensuring both content completeness and faithful visual presentation.

\noindent Collectively, these examples across both template-free and template-based settings show that AutoPage produces more faithful and reliable webpages than the baseline, maintaining both structural consistency and content integrity in challenging cases.

\section{Effectiveness of Human-in-the-loop Feeback}
\label{sec:appendix-human-in-the-loop}
\begin{figure*}[!t]
            \centering
            \includegraphics[width=0.85\textwidth]{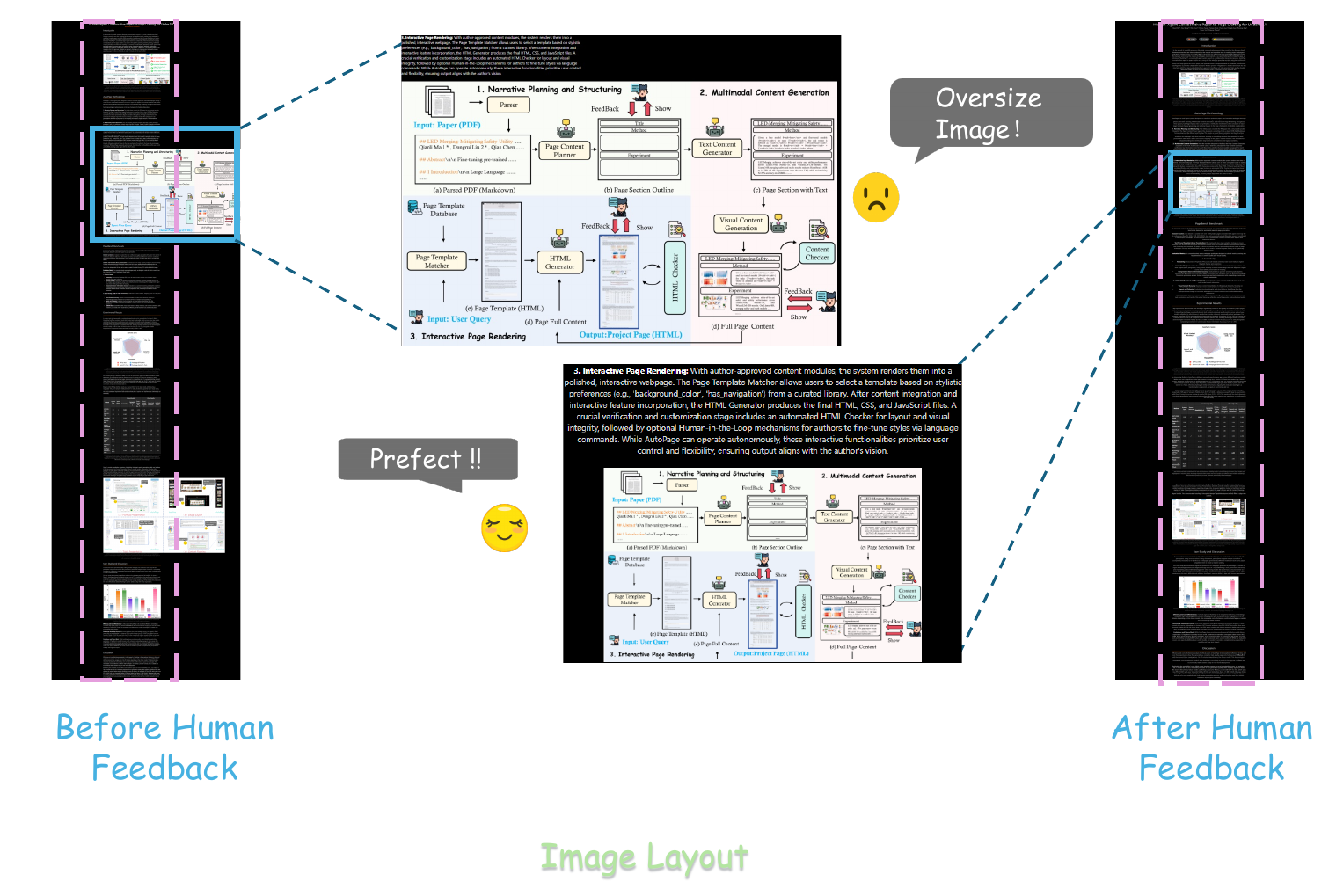}
            \caption{\textbf{Correcting Page Layout with Human Feedback.} The baseline generation (left) results in a poor layout with an oversized image. By integrating human feedback, AutoPage (right) produces a corrected layout with a properly-sized image.}
     \label{fig:app-human1}
    \vspace{-10pt}
\end{figure*}

\begin{figure*}[!t]
            \centering
            \includegraphics[width=0.85\textwidth]{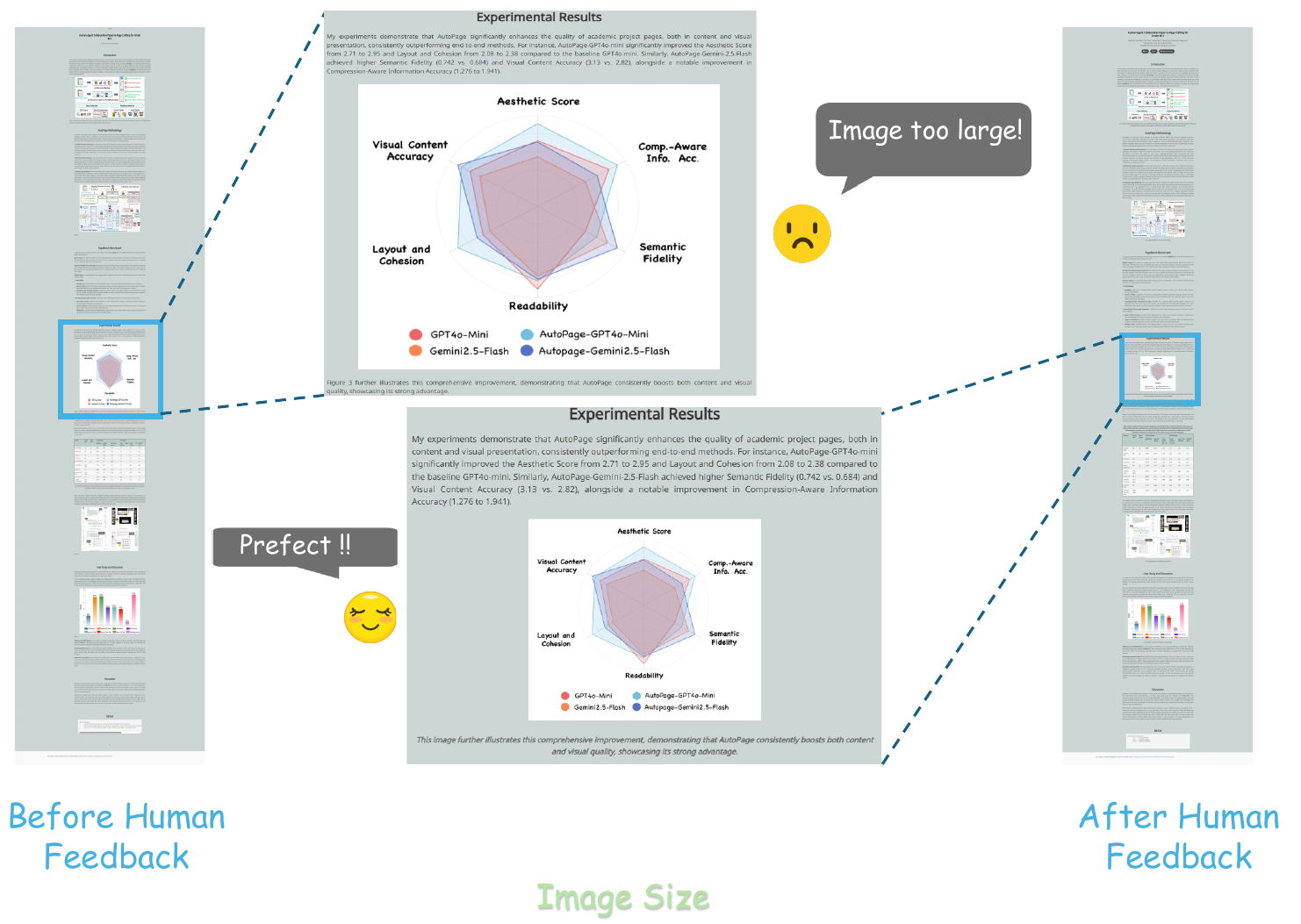}
            \caption{\textbf{Impact of Human Feedback on Visual Layout.} The initial page generated without human feedback (left) suffers from a flawed layout, featuring a disproportionately large image. In contrast, after incorporating human feedback, AutoPage (right) corrects the layout and renders the image at an appropriate size.}
     \label{fig:app-human2}
    \vspace{-10pt}
\end{figure*}

\begin{figure*}[!t]
            \centering
            \includegraphics[width=0.85\textwidth]{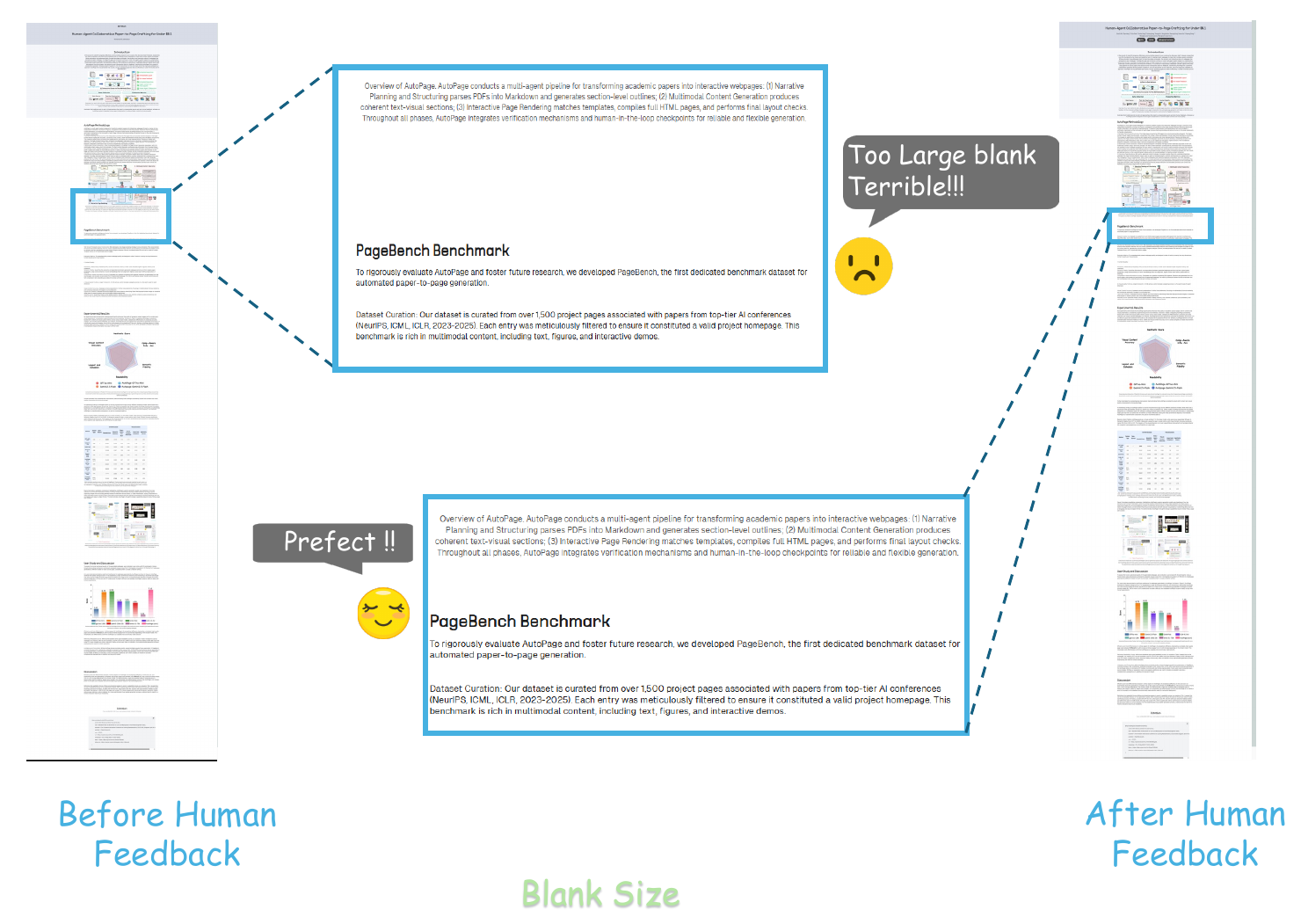}
            \caption{\textbf{Impact of Human Feedback on Vertical Layout Spacing.} The initial page generated by AutoPage without human feedback (left) exhibits excessive vertical whitespace between content modules, resulting in a sparse and poorly structured layout. In contrast, after incorporating human feedback, the system (right) corrects the spacing to produce a more compact and visually coherent page.}
     \label{fig:app-human3}
    \vspace{-10pt}
\end{figure*}

\begin{figure*}[!t]
            \centering
            \includegraphics[width=0.85\textwidth]{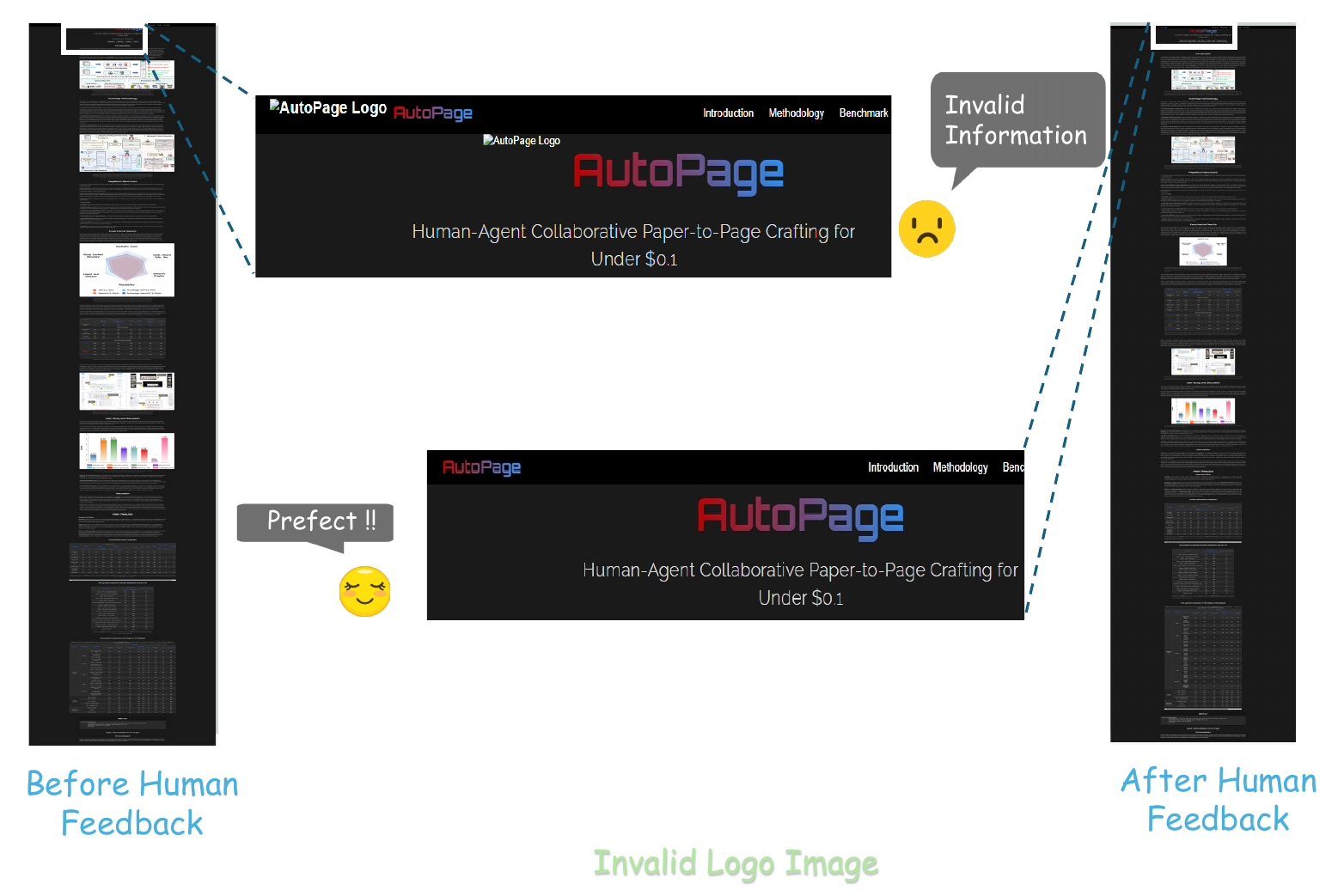}
            \caption{\textbf{Human-in-the-Loop Correction for Page Assets.} This figure demonstrates how human feedback is used to refine UI components. The initial output (left) features an incorrect or broken logo image in the header. After processing the feedback, AutoPage generates a corrected version (right) where the erroneous asset has been removed.}
     \label{fig:app-human4}
    \vspace{-10pt}
\end{figure*}

In this section, we demonstrate the effectiveness of human feedback in AutoPage, shown in Fig.~\ref{fig:app-human1}, Fig.~\ref{fig:app-human2}, Fig.~\ref{fig:app-human3}, Fig.~\ref{fig:app-human4}. 

As demonstrated in Fig.~\ref{fig:app-human1} and Fig.~\ref{fig:app-human2}, the initial automated generation can produce suboptimal layouts, such as those with disproportionately large images that disrupt the page structure. With guidance from human feedback, AutoPage effectively corrects these scaling issues to restore a balanced and visually appropriate layout. Fig.~\ref{fig:app-human3} highlights a different type of layout refinement, where human intervention resolves excessive vertical whitespace between content modules, leading to a more compact and visually coherent page. Beyond layout adjustments, Fig.~\ref{fig:app-human4} illustrates how the feedback mechanism can rectify content-level errors by identifying and removing erroneous assets, such as an incorrect logo in the header.

\begin{table}[h!]
\centering
\caption{\textbf{Effect of Human-in-the-Loop (HITL) visual feedback.} Participants were allowed to provide one round of high-level page-level visual feedback. Results show that even minimal HITL leads to clear improvements, especially in structural layout quality.}
\vspace{-10pt}
\label{tab:hitl_effect}
\resizebox{\linewidth}{!}{
\begin{tabular}{l cc}
\toprule
\textbf{Metric} & \textbf{w/o HITL} & \textbf{w/ HITL} \\
\midrule
\multicolumn{3}{l}{\textit{Visual Quality}} \\
\quad Visual Accuracy $\uparrow$            & 3.13 & \textbf{3.15} \\
\quad Layout \& Cohesion $\uparrow$         & 2.15 & \textbf{3.10} \\
\quad Aesthetic Score $\uparrow$            & 2.69 & \textbf{2.75} \\
\bottomrule
\end{tabular}
}
\vspace{-10pt}
\end{table}

We conducted an additional controlled comparison where participants provided a \textbf{single round of page-level visual feedback}, after which the webpage was regenerated accordingly. We focused on visual feedback because users find structural and layout issues far easier and more immediate to comment on than detailed textual corrections. As shown in Tab.~\ref{tab:hitl_effect}, we observe a substantial improvement in \textbf{layout and cohesion}, while content accuracy and aesthetics increase moderately. This pattern reflects how human-in-the-loop is typically used: participants mostly commented on structural issues, including section ordering, spacing, figure placement, and alignment. These issues are easy to identify and correct in a single feedback round. In contrast, deeper content or fine-grained stylistic refinements often require multiple iterations, which were beyond this single-turn setup. These findings show that even minimal human input produces clear, measurable gains in structural visual quality, confirming the practical utility of the human-in-the-loop mechanism without requiring heavy human effort.

Collectively, these examples and experiments confirm the \textit{versatility and precision of the human-in-the-loop process}, enabling fine-grained corrections that range from layout spacing and image scaling to content validation, thereby significantly enhancing the final output quality.

\section{Prompt Templates}
\label{sec:appendix-prompt-templates}
In this section, we present the prompt templates we used for each component in AutoPage.

\onecolumn
\begin{tcolorbox}[filter_figures]

  \textbf{System Prompt}

You are a helpful academic expert. You need to determine which section of the paper each image and table in the figures belongs to from given research paper's contents and figures.

\textbf{Template Description}

Below is the figures with descriptions, paths, width and height in the paper:

\begin{verbatim}
<figures>
{{figures}}
</figures>
\end{verbatim}

I have already generated the text-based project page content as follows:

\begin{verbatim}
<project_page_content>
{{project_page_content}}
</project_page_content>
\end{verbatim}

The paper content is as follows:

\begin{verbatim}
<paper_content>
{{paper_content}}
</paper_content>
\end{verbatim}

\textbf{Tasks}

\begin{enumerate}
    \item Determine which section of the article each image and table in the figures belongs to, and then add a field called \texttt{original\_section} to every figure in the original figures, filling it with the determined section. If a figure does not appear in the paper content, then \texttt{original\_section} should be set to null. Your output should be json format.
    \item Extract figure and table tags from figure or table captions. Key of these tags is \texttt{tag}.
    \item Remove the extracted tag from caption of each figure.
\end{enumerate}

\textbf{Output Format}

\begin{verbatim}
```json
<Your output>
```
\end{verbatim}

\end{tcolorbox}

\begin{tcolorbox}[section_generation]

    \textbf{System Prompt}

    You are an expert content planner specializing in creating engaging project pages for research papers. Your role is to analyze research content and plan an effective structure that communicates the research clearly and professionally.
    
    You will be given:
    \begin{enumerate}
        \item Research Paper in markdown format.
        \item List of images extracted from the paper.
        \item List of tables extracted from the paper.
    \end{enumerate}
    
    Your goal is to create a section plan that organizes the research into an effective project page structure.
    
    \textbf{Template Description}
    
    Please analyze the paper content and identify the key sections that should be included in the project page. For each section, provide a concise description of what should be included. First, determine the paper type:
    
    \begin{itemize}
        \item \textbf{For methodology research papers:} Focus on method description, experimental results, and research methodology.
        \item \textbf{For benchmark papers:} Highlight task definitions, dataset construction, and evaluation outcomes.
        \item \textbf{For survey/review papers:} Emphasize field significance, key developmental milestones, critical theories/techniques, current challenges, and emerging trends.
    \end{itemize}
    
    Note that the specific section names should be derived from the paper's content. Related sections can be combined to avoid fragmentation. Limit the total number of sections to maintain clarity.
    
    You must include some section that describe the methodology and experiments.
    
    Do not include acknowledgments or references sections. Do not include related work and experiment setting sections.
    
    The number of sections you design must not exceed five.
    
    Return the result as a flat JSON object with section names as keys and descriptions as values, without nested structures. You MUST use Markdown code block syntax with the json language specifier.
    
    \textbf{Paper Content:}
    
    \begin{verbatim}
    {{paper_content}}
    \end{verbatim}
    
    \textbf{Output Format}
    
    \begin{verbatim}
    ```json
    <Your generated section json content>
    ```
    \end{verbatim}
\end{tcolorbox}

\begin{tcolorbox}[text_content_generation]

  \textbf{System Prompt}

You are a helpful academic expert and web developer, who is specialized in generating a text-based paper project page, from given contents.

\textbf{Template Description}

You will be given the research paper's paper markdown content, figures, and a section plan that describe what content should be included in each section of the project page.

Your task is to fill in the actual content for each section based on the requirements outlined in the section plan and the content of the research paper.

In the project page, you should introduce it from the author's perspective rather than from a third-party viewpoint. This content will ultimately be displayed on the project page. The content you generated must include all key components of the paper.

Below is the image and table figures with descriptions and paths in the paper:

\begin{verbatim}
<figures>
{{figures}}
</figures>
\end{verbatim}

Below is the content of the paper:

\begin{verbatim}
<paper_content>
{{paper_content}}
</paper_content>
\end{verbatim}

Section Plan:

\begin{verbatim}
{{format_instructions}}
\end{verbatim}

If figures can effectively convey the poster content, simplify the related text to avoid redundancy. Include essential mathematical formulas where they enhance understanding.

Don't leave any important content in the research paper. If the paper content has a conclusion section, this section should not contain any figures.

\textbf{Do not include any tag of figure or table in the text}

Your output must be in JSON format, and the section names in your output must exactly match those in the section plan.

\textbf{Output Format}

\begin{verbatim}
```json
<Your output>
  \end{verbatim}
Ensure all sections are precise, concise.

\end{tcolorbox}

\begin{tcolorbox}[full_content_generation]

  \textbf{System Prompt}

You are a helpful academic expert and web developer, who is specialized in generating a paper project page, from given research paper's contents and figures.

\textbf{Template Description}

Below is the figures with descriptions, paths, width and height in the paper:

\begin{verbatim}
<figures>
{{figures}}
</figures>
\end{verbatim}

I have already generated the text-based project page content as follows:

\begin{verbatim}
<project_page_content>
{{project_page_content}}
</project_page_content>
\end{verbatim}

The paper content is as follows:

\begin{verbatim}
<paper_content>
{{paper_content}}
</paper_content>
\end{verbatim}

\textbf{Task Requirements}

Your task is inserting figures into the project page content using figure index notation as:

\begin{verbatim}
![figure_description][figure_path][width=figure_width, height=figure_height]
\end{verbatim}
\begin{verbatim}(figure_index)
\end{verbatim}
For example:

\begin{verbatim}
![Overview]["assets/paper-picture-8.png"][width=1068, height=128](8)
\end{verbatim}

\textbf{Specific Requirements}

\begin{enumerate}
    \item The \texttt{figure\_index} MUST be an integer starting from 1, and no other text should be used in the figure\_index position.
    \item Each figure should be used at most once, with precise and accurate placement.
    \item Prioritize pictures and tables based on their relevance and importance to the content.
    \item The teaser figure that appears early in the paper must be included in the content.
    \item Don't leave any important figure in the research paper.
    \item If a chapter has multiple tables, \textbf{only the one most relevant} to the chapter should be included.
    \item Your output must be in JSON format, and the section names in your output must exactly match those in the \texttt{project\_page\_content}.
    \item Please ensure that the images you insert are closely related to the context and align well with the content of the section.
\end{enumerate}

\textbf{Output Format}

\begin{verbatim}
```json
<Your output>
```
\end{verbatim}
  
\end{tcolorbox}

\begin{tcolorbox}[html_generation]
\textbf{System Prompt}

You are an expert web developer specializing in creating professional project pages for research papers. You have extensive experience in HTML5, CSS3, responsive design, and academic content presentation. Your goal is to create a complete, production-ready HTML project page that is visually appealing and professional.

\textbf{Template Description}

Instructions:

Generate a complete, production-ready HTML project page based on the provided project page content and html template.

Project Page Content:

\begin{verbatim}
{{ generated_content }}
\end{verbatim}

HTML Template:

\begin{verbatim}
{{ html_template }}
\end{verbatim}

\textbf{Requirements}

\begin{enumerate}
    \item The HTML files you generate should follow the format and style of the reference HTML template as closely as possible, but you can replace the content in the reference HTML template.
    \item The content of your project page should be filled completely with the project page content, without any omissions.
    \item All content sections in Project Page Content should be properly formatted in your html file.
    \item Images and tables in Project Page Content should be integrated into your html using the correct image path or table path.
    \item You should make sure that paths of css files included in the html file should not be changed.
    \item If there is a formula in the generated content, please add the relevant js code such as:
    
\begin{verbatim}
<script>
 window.MathJax = {
 tex: {
 inlineMath: [['$', '$'], ['\\(', '\\)']],
 displayMath: [['$$', '$$'], ['\\[', '\\]']]
 }
 };
</script>
<script src="https://cdn.jsdelivr.net/npm/mathjax@3/es5/tex-mml-chtml.js">
</script>
\end{verbatim}

    \item For images and tables, only the caption needs to be retained and the tag before the caption needs to be deleted (e.g. Figure 1., Table 2.).
    \item Do not place the table in a small container. If the table is large, place it in a larger container.
    \item \textbf{The image (not table) should be placed in the container to limit its size}.
    \item Formulas should not be placed in separate paragraphs.
    \item \textbf{Be careful not to let the image overflow the screen, the screen width is generally 1280. You can add a container that is the same width as the screen to limit}.
\end{enumerate}

\textbf{Layout Specification}

\begin{itemize}
    \item For example, if there are two images in two columns whose aspect ratios are 1.2 and 2 respectively, the flex grow of two columns should be 1.2 and 2 respectively, to make the columns have the same height.
    \item Calculate the size of each image based on column width and aspect ratios. Add comment \texttt{<!-- width = display\_width, height = display\_height -->} before each image.
    \item Rearrange the structure and order of sections, texts and images to make the height of each column in the same group approximately the same.
    \item For example, if there are too many images in one section that make the height of the column too large, group the images into columns.
    \item The display width of each image should not be too large or too small compared to its original width.
    \item In each section, if an image or a table is placed in a single column layout, it should be horizontally centered within that column or content block (not the full page). Use CSS techniques such as \texttt{display: block; margin: auto;} to achieve proper visual centering.
    \item There should be a certain amount of spacing between adjacent images.
    \item All sections on the entire project page must be in a single column and span the full width of the page.
    \item Formulas do not need to be opened in a separate paragraph such as section-text block (such as \texttt{<p class="section-text" style="text-align: center;">}).
\end{itemize}

\textbf{Output Format}

\begin{verbatim}
```html
<html_content>
```
\end{verbatim}
\end{tcolorbox}

\begin{tcolorbox}[full_content_review]
\textbf{System Prompt}

You are an expert reviewer for scientific project pages. Your task is to carefully \textbf{review the generated content} by comparing it with the original paper content and figures.

You will be given three inputs:

\begin{enumerate}
    \item \textbf{paper\_content}: the original scientific content of the paper.
    \item \textbf{figures}: the list of figures (including captions, tag, intended placement, and meaning).
    \item \textbf{generated\_content}: the project page content automatically generated by another agent.
\end{enumerate}

You cannot violate these basic rules below for the project page when generating suggestions.

\begin{enumerate}
    \item \textbf{Number of tables in the whole page must be less than or equal to 3.}
    \item \textbf{Any figure or table can just appear once in the content.}
    \item \textbf{Include at least one table in experiment and ablation section if these two are included in generated sections.}
    \item \textbf{Include at least one image in visualization section if it is included in generated sections.}
\end{enumerate}

\textbf{Remember that you should just restrict number of tables under 4, rather than restrict the total number of visual elements in the whole content.}

You can know if a visual element is a table by its tag.

\textbf{You should first get the number of tables and number of figures respectively in the content and then tell if the number of tables is more than 3.}

Your review must focus on the following dimensions:

\textbf{1. Figure Placement and Usage}

\begin{itemize}
    \item Verify whether figures are inserted in the correct sections according to their meaning in the paper.
    \item For each section, you should check whether the text content and captions of figures and tables it includes is tightly related.
    \item Check if two figures convey similar idea. If it is, you should remain the more important figure.
\end{itemize}

\textbf{2. Relation between figures and text}

\begin{itemize}
    \item Check if the core idea that a figure shows is mentioned in text of its section.
    \item If the correlation between them is weak, please suggest to remove the figure or move it to other section.
\end{itemize}

\textbf{3. Number of tables}

\begin{itemize}
    \item You should tell whether the number of tables is more than 2. If it is, you \textbf{should choose 2 most important table to remain}.
    \item Do not restrict the number of figures.
\end{itemize}

Below is the figures with captions, paths, width and height in the paper:

\begin{verbatim}
<figures>
{{figures}}
</figures>
\end{verbatim}

Below is the tables with captions, paths, width and height in the paper:

\begin{verbatim}
<tables>
{{tables}}
</tables>
\end{verbatim}

The paper content is as follows:

\begin{verbatim}
<paper_content>
{{paper_content}}
</paper_content>
\end{verbatim}

The generated project page content is as follows:

\begin{verbatim}
<project_page_content>
{{generated_content}}
</project_page_content>
\end{verbatim}

\textbf{Requirements}

\begin{enumerate}
    \item Do not suggest adding or deleting entire sections.
    \item The generated project page content should present the more important parts of the paper content in a concise manner, so your review should not require including too many unimportant details.
    \item Remember that the original section of a image is not necessary to be same as the section it belongs to in the page. Do not correlate the two sections together.
    \item Do not give suggestions of including figure Captions, because they will be included during the generation of html, not full content.
    \item Do not give suggestion to change any text content in any section, you can just suggest to add or delete or move figures and tables.
    \item Tables and Figures from Ablation section in the paper content should belong to Experiment section in the generated content if Ablation is not included in generated sections.
\end{enumerate}

\textbf{Output Format}

You must return your review in \textbf{strict JSON format} with the following fields:

\textbf{Output Format}

You must return your review in \textbf{strict JSON format} with the following fields:

\begin{verbatim}
{
  "weakness": [
    "weakness_1",
    "weakness_2"
  ],
  "strength": [
    "strength_1",
    "strength_2"
  ],
  "suggestion": [
    "suggestion_1",
    "suggestion_2"
  ]
}
\end{verbatim}

\end{tcolorbox}
\begin{tcolorbox}[full_content_revise]
\textbf{System Prompt}

Please \textbf{revise the previously generated project page content} according to the review below:

\begin{verbatim}
<review_content>
{{review_content}}
</review_content>
\end{verbatim}

\textbf{Instructions}

\begin{enumerate}
    \item Carefully read the \texttt{weakness}, \texttt{strength}, and \texttt{suggestion} fields in the review JSON.
    \item Improve the previously generated content by:
    \begin{itemize}
        \item Fixing weaknesses
        \item Preserving strengths
        \item Applying suggestions directly and concretely
    \end{itemize}
    \item Ensure the revised content is:
    \begin{itemize}
        \item \textbf{Accurate} (aligned with the original intent of the paper and figures).
        \item \textbf{Clear and fluent} (scientifically precise, grammatically correct, and concise).
        \item \textbf{Well-structured} (logical flow, correct figure placement).
    \end{itemize}
    \item Please do not add or remove any sections.
    \item Do not change the name of any section in the page content.
    \item Do not include two identical figures in the page content.
    \item Do not change any text content.
\end{enumerate}

\textbf{Output Format}

\begin{verbatim}
```json
<Your output>
```
\end{verbatim}

\end{tcolorbox}

\begin{tcolorbox}[html_review]
\textbf{System Prompt}

You are a professional reviewer specializing in images and figures on research project pages. Your task is to inspect \textbf{all images individually and meticulously}, considering \textbf{only width, vertical spacing, and internal text size}. Return a \textbf{summary report} that aggregates feedback across all images.

\textbf{Evaluation Criteria}

\textbf{1. Width of images}

\begin{itemize}
    \item Judge strictly whether each image is too wide, too narrow, or within the ideal visual range (approximately 70–90\% of the main text block).
    \item Mark as a \textbf{weakness} if:
    \begin{itemize}
        \item The image exceeds the main text block width.
        \item The image is clearly smaller than 65\% of the text block width.
    \end{itemize}
    \item \textbf{Allowed adjustment:} specify a \textbf{percentage of the original image width} (e.g., ``shrink to 90\% of original width'').
    \item \textbf{Special rule for containers:}
    \begin{itemize}
        \item \textbf{Only if an image clearly overflows outside the viewport/screen (not just wider than text block), recommend adding a container restriction.}
        \item In all other cases, simply suggest proportional resizing by percentage.
    \end{itemize}
\end{itemize}

\textbf{2. Vertical spacing}

\begin{itemize}
    \item Check if the spacing above and below each image is visually balanced.
    \item Mark as a \textbf{weakness} if one side is noticeably larger than the other.
    \item \textbf{Allowed adjustment:} specify exact pixel values (e.g., ``set both top and bottom margin to 24px'').
    \item Spacing must be consistent with surrounding blocks to maintain overall page rhythm.
\end{itemize}

\textbf{3. Text inside images}

\begin{itemize}
    \item Judge whether text inside the image is \textbf{disproportionately larger than body text (\textasciitilde 12–14px)}.
    \item \textbf{Font inside images cannot be changed directly.}
    \item \textbf{Allowed adjustment:} suggest resizing the \textbf{whole image} by a specific percentage of its original width (e.g., ``shrink to 80\% of original width'') to bring internal text visually closer to body text size.
\end{itemize}

\textbf{4. Strictness Rules}

\begin{itemize}
    \item Only consider the three aspects above: width, vertical spacing, internal text.
    \item Skip tables when checking internal text size.
    \item Provide precise, actionable suggestions using percentages for width adjustments and pixels for margins.
    \item Be conservative: report all issues, even if minor, with clear reference to the affected image.
    \item \textbf{Container restriction is suggested only if the image overflows outside the viewport.}
    \item Formulas should not be placed in separate paragraphs.
\end{itemize}

\textbf{Output Format}

Return a \textbf{single JSON object} with three arrays that summarize all images:

\begin{itemize}
    \item \texttt{"weaknesses"}: list all weaknesses across all images, each item must indicate the image it comes from and the specific problem.
    \item \texttt{"strengths"}: list all positive aspects across all images, each item must indicate the image it comes from.
    \item \texttt{"suggestions"}: list all actionable suggestions for all images, using \textbf{specific percentages for width adjustments} and \textbf{pixels for vertical spacing}.
    \item \textbf{Only include ``add container restriction'' if the image actually overflows the viewport.}
\end{itemize}

\begin{verbatim}
```json
{
  "weaknesses": [
    "weakness_1",
    "weakness_2"
  ],
  "strengths": [
    "strength_1",
    "strength_2"
  ],
  "suggestions": [
    "suggestion_1",
    "suggestion_2"
  ]
}
\end{verbatim}
\end{tcolorbox}
\begin{tcolorbox}[html_revise]
\textbf{System Prompt}

You are an expert web developer specializing in creating professional project pages for research papers. You have extensive experience in HTML5, CSS3, responsive design, and academic content presentation. Your goal is to produce a complete, production-ready HTML project page that is visually appealing, professional, and adheres to specified constraints.

\textbf{Template Description}

Instructions:

Generate a refined, production-ready HTML project page based on the \textbf{existing HTML} and the provided \textbf{suggestions}.

Existing HTML:

\begin{verbatim}
{{ existing_html }}
\end{verbatim}

Suggestions:

\begin{verbatim}
{{ suggestions }}
\end{verbatim}

\textbf{Requirements}

\begin{enumerate}
    \item Apply all the suggestions carefully to the existing HTML without omitting any improvements.
    \item Preserve the original formatting, style, and constraints from the existing HTML, unless explicitly adjusted by the suggestions.
    \item All content sections must remain properly formatted and intact; do not remove or lose any original content.
    \item Images and tables should retain correct paths and aspect ratios; apply size adjustments or centering only if suggested.
    \item Maintain responsive design, single-column full-width layout, and professional visual presentation.
    \item Ensure proper spacing, alignment, symmetry, and column height balance as specified in the previous layout rules.
    \item Comment \texttt{<!-- width = display\_width, height = display\_height -->} before each image whose size is adjusted according to column width and aspect ratio.
    \item Center images or tables horizontally within their column or content block where applicable.
    \item All other previous layout constraints and formatting rules must be respected.
    \item Modify the image size according to the suggestions, making sure it is centered and there should be a certain amount of space between the images.
\end{enumerate}

\textbf{Output Format}

\begin{verbatim}
```html
<html_content>
```
\end{verbatim}
\end{tcolorbox}

\begin{tcolorbox}[vlm_aesthetics_judge]
\textbf{System Prompt:}
  You are an extremely strict visual aesthetics reviewer. Focus solely on the overall aesthetic feel of the page or project presentation, including color scheme, style consistency, visual hierarchy, text-image coordination, and overall visual impression.
  Do not consider the accuracy of formulas or specific content of text or images, only evaluate aesthetics and visual feel. Do not easily give high scores unless the overall aesthetics reach an extremely high level.

\textbf{Template:} 

Scoring Description:\\
    Five-point scale:\\
    \textbf{1~point}:
        \begin{itemize}
            \item The overall color scheme of the page or content is chaotic or conflicting, resulting in a very poor visual experience.
            \item The style is inconsistent, with no sense of hierarchy among elements, and the overall impression is cluttered.
            \item Poor coordination between images, text, or charts, causing visual fatigue.
        \end{itemize}
    \textbf{2~points} 
        \begin{itemize}
            \item The color scheme or style shows obvious inconsistencies, with some areas having weak aesthetic appeal.
            \item The hierarchy of elements is slightly chaotic, with unclear visual guidance.
            \item Text-image coordination is average, and the overall impression is slightly cluttered.
        \end{itemize}
    \textbf{3~points} 
        \begin{itemize}
            \item The color scheme is generally reasonable but has minor conflicts or imbalances.
            \item The style is relatively unified, but some elements are slightly uncoordinated.
            \item Text-image coordination is good, but there is still room for improvement in overall aesthetics.
        \end{itemize}
    \textbf{4~points} 
        \begin{itemize}
            \item The color scheme is comfortable, with consistent style and good overall visual aesthetics.
            \item The hierarchy of elements is clear, with reasonable visual guidance and high text-image coordination.
            \item  The overall impression is comfortable, with strong visual appeal.
        \end{itemize}
     \textbf{5~points} 
        \begin{itemize}
            \item Rarely used; reserved for publication-level visual aesthetics.
            \item Color, style, hierarchy, and layout are perfectly unified, with harmonious and beautiful overall visuals.
            \item   Text-image coordination is exceptional, with natural visual rhythm, an excellent impression, and no flaws.
        \end{itemize}
  
  - Example Output:
  \begin{verbatim}
  {
    "reason": "xx",
    "score": int
  }
  \end{verbatim}
     Please provide scores strictly and conservatively.

\end{tcolorbox}

\begin{tcolorbox}[vlm_element_judge]
\textbf{System Prompt:}
  You are an extremely strict visual elements reviewer. Focus solely on the presentation of formulas and images, as well as the relevance of images to paragraph content.
  Do not consider webpage layout, paragraph formatting, or overall design. Do not easily give high scores unless the visual elements fully meet the highest standards.

\textbf{Template:} 
Scoring Description:\\
    Five-point scale:\\
    \textbf{1~point}:
        \begin{itemize}
            \item Formulas are incompletely displayed or entirely missing.
            \item Images are almost irrelevant to paragraph content or entirely missing.
            \item Colors are hard to distinguish, affecting comprehension.
        \end{itemize}
    \textbf{2~points} 
        \begin{itemize}
            \item Some formulas or images are displayed correctly, but others are missing or incorrect.
            \item Image labels or captions are unclear, with weak relevance to text.
            \item Colors or clarity have some issues, making comprehension difficult.
        \end{itemize}
    \textbf{3~points} 
        \begin{itemize}
            \item Most areas have a reasonable layout, but there are occasional issues with uneven white space or slightly oversized/undersized images.
            \item The page is generally visually balanced but has slight inconsistencies.
            \item Minor impact on reading or comprehension, but overall acceptable.
        \end{itemize}
    \textbf{4~points} 
        \begin{itemize}
            \item The page layout is good, with well-proportioned visual elements.
            \item White space is reasonable, and image sizes are appropriate.
            \item The overall visual experience is comfortable, with smooth information delivery.
        \end{itemize}
     \textbf{5~points} 
        \begin{itemize}
            \item Rarely used; reserved for publication-level page layout.
            \item White space and image sizes are perfectly balanced, with a natural visual rhythm.
            \item   The page is visually harmonious, offering an excellent reading experience with no distractions.
        \end{itemize}
  
  - Example Output:
  \begin{verbatim}
  {
    "reason": "xx",
    "score": int
  }
  \end{verbatim}
     Please provide scores strictly and conservatively.

\end{tcolorbox}

\begin{tcolorbox}[vlm_layout_judge]
\textbf{System Prompt:}
  You are an extremely strict webpage layout reviewer. Focus solely on the visual layout and typesetting experience of the page, without considering the clarity of formulas, images, or their relevance to text. Pay attention to issues such as white space, image size, and visual balance, particularly noting whether large areas of white space or oversized images disrupt the visual effect. Do not easily give high scores unless the layout is highly reasonable.

\textbf{Template:} 
Scoring Description:\\
    Five-point scale:\\
    \textbf{1~point}:
        \begin{itemize}
            \item The page layout is extremely chaotic or cluttered.
            \item Large areas of white space or oversized images disrupt the overall visual effect.
            \item The page is visually unbalanced, resulting in a poor reading experience.
        \end{itemize}
    \textbf{2~points} 
        \begin{itemize}
            \item Some areas have a reasonable layout, but there are still noticeable large areas of white space or oversized images.
            \item The page lacks visual balance, with a mediocre overall impression.
            \item Some element arrangements may hinder information acquisition.
        \end{itemize}
    \textbf{3~points} 
        \begin{itemize}
            \item Most formulas and images are displayed correctly, but there are noticeable issues with clarity, style, or annotations.
            \item Some images have average relevance to paragraph content.
            \item Minor issues with labels or colors, but they do not significantly affect comprehension.
        \end{itemize}
    \textbf{4~points} 
        \begin{itemize}
            \item Formulas are fully displayed, and images are clear and highly relevant to paragraph content.
            \item Labels, legends, and colors are reasonable and aid comprehension.
            \item Style is relatively consistent, with overall good visual presentation.
        \end{itemize}
     \textbf{5~points} 
        \begin{itemize}
            \item Rarely used; reserved for publication-level visual presentation.
            \item Formulas are perfectly displayed, and images are clear and highly relevant to text.
            \item   Labels and legends are flawless, with unified colors and style, and impeccable visual presentation.
        \end{itemize}
  
  - Example Output:
  \begin{verbatim}
  {
    "reason": "xx",
    "score": int
  }
  \end{verbatim}
   Please provide scores strictly and conservatively.

\end{tcolorbox}
\twocolumn

\end{document}